\documentclass[%
reprint,
superscriptaddress,
nofootinbib,
 amsmath,amssymb,
 aps,
prb,
floatfix,
]{revtex4-2}

\usepackage{graphicx}% Include figure files
\usepackage{dcolumn}% Align table columns on decimal point
\usepackage{bm}% bold math
\usepackage{braket}
\usepackage{physics}
\usepackage{comment}
\usepackage{color}
\usepackage{txfonts}
\usepackage{{mhchem}}
\usepackage{bbm}
\usepackage[colorlinks=true,urlcolor=blue,citecolor=blue,linkcolor=blue,breaklinks=true]{hyperref}

\usepackage[whole]{bxcjkjatype} 

\begin{document}

%\preprint{APS/123-QED}

\title{Thermal Hall responses in frustrated honeycomb spin systems}

\author{Kosuke Fujiwara}
\affiliation{Department of Applied Physics, The University of Tokyo, Hongo, Tokyo, 113-8656, Japan}

\author{Sota~Kitamura}
\affiliation{Department of Applied Physics, The University of Tokyo, Hongo, Tokyo, 113-8656, Japan}

\author{Takahiro~Morimoto}
\affiliation{Department of Applied Physics, The University of Tokyo, Hongo, Tokyo, 113-8656, Japan}
\affiliation{JST, PRESTO, Kawaguchi, Saitama, 332-0012, Japan}

\date{\today}% It is always \today, today,

\begin{abstract}
We study geometrical responses of magnons driven by a temperature gradient in frustrated spin systems. While Dzyaloshinskii-Moriya (DM) interactions are usually incorporated to obtain geometrically nontrivial magnon bands, here we investigate thermal Hall responses of magnons that do no rely on the DM interactions.
Specifically, we focus on frustrated spin systems with sublattice degrees of freedom and show that a nonzero Berry curvature requires breaking of an effective $PT$ symmetry.
According to this symmetry consideration, we study the $J_1$-$J_2$-$J_2^\prime$ Heisenberg models on a honeycomb lattice as a representative example, and demonstrate that magnons in the spiral phase support the thermal Hall effect once we introduce a magnetic field and asymmetry between the two sublattices.
We also show that driving the magnons by a temperature gradient induces spin current generation (i.e., magnon spin Nernst effect) in the $J_1$-$J_2$-$J_2^\prime$ Heisenberg models.

\end{abstract}

\maketitle

%\tableofcontents

\section{\label{sec:level1}Introduction \protect}

A magnon is an elementary excitation of spin waves in magnetic materials. Magnon transport is attracting growing interests in both fundamental and technological aspects~\cite{Chumak2015MagnonSpintronics}.
For example, magnons can transfer spins without Joule heating and are expected to play an essential role in spintronics as a platform for low energy consumption devices.
In particular, antiferromagnetic spintronics is attracting a keen attention because antiferromagnets have no leakage magnetic field in contrast to conventional ferromagnets~\cite{Jungwirth2016AntiferromagneticSpintronics,Baltz2018AntiferromagneticSpintronics}.

Since magnons are charge neutral quasiparticles, they cannot be directly driven by electric fields, unlike electrons. 
Instead, a temperature gradient can induce a magnon flow,
which leads to various thermal responses in magnets, including 
the spin Seebeck effect~\cite{Xiao2010TheoryEffect}, the magnon spin Nernst effect~\cite{Zyuzin2016MagnonAntiferromagnets,Cheng2016SpinAntiferromagnets,Shiomi2017ExperimentalMnPS3}, and the thermal Hall effect~\cite{Katsura2010TheoryMagnets,Onose2010ObservationEffect}.
The thermal Hall effect and the magnon spin Nernst effect are of particular interest because they are related to a nontrivial geometry of the magnon bands through the Berry curvature ~\cite{Matsumoto2011RotationalEffect,Matsumoto2011TheoreticalFerromagnets,Matsumoto2014ThermalInteraction, Zyuzin2016MagnonAntiferromagnets,Cheng2016SpinAntiferromagnets,Zhang2018Spin-NernstInsulator}.

Most previous studies on such geometrical thermal responses of magnons rely on Dzyaloshinskii-Moriya (DM) interactions to obtain geometrically nontrivial 
magnon bands with nonzero Berry curvature.
For example, the thermal Hall effect has been studied in an antiferromagnetic Heisenberg model with DM interactions on a kagome lattice~\cite{Park2019TopologicalAntiferromagnets,Laurell2018MagnonInteractions,Doki2018SpinAntiferromagnet,Mook2019ThermalAntiferromagnets,Owerre2017TopologicalAntiferromagnets}
and a honeycomb lattice~\cite{Owerre2017NoncollinearInsulator}.
Similarly, the magnon spin Nernst effect has been studied in a  Heisenberg model with a DM interaction~\cite{Zyuzin2016MagnonAntiferromagnets,Cheng2016SpinAntiferromagnets,Zhang2018Spin-NernstInsulator}.

The DM interaction introduces a nontrivial geometry to magnon bands in two fashions. First, the DM interaction acts as a virtual magnetic field for magnons, leading to non-zero Berry curvature~\cite{Katsura2010TheoryMagnets,Ideue2012EffectInsulators}. In this case, there exists a condition for the lattice geometry to support nonzero Berry curvature because edge shared lattices results in cancellation of such virtual magnetic field between the neighboring plaquettes. For example, a kagome lattice supports a thermal Hall response with this mechanism. 
Second, the DM interaction can also introduce an effective non-Abelian gauge field for magnons with multiple internal degrees of freedom. In particular, when we consider a bipartite lattice with $AB$ sublattices, the DM interaction can behave as an SU(2) gauge field for the sublattice degree of freedom. This mechanism is advantageous over the first one in that the lattice geometry is not restricted~\cite{Kawano2019ThermalAntiferromagnet}.
In both cases, however, the DM interaction is usually small except for a few limited systems because the DM interaction originates from the spin-orbit interaction \cite{Dzyaloshinsky1958AAntiferromagnetics,Moriya1960AnisotropicFerromagnetism}. Therefore, geometrically nontrivial magnon bands that do not rely on the presence of the DM interaction are desired for an enhancement of thermal Hall responses in magnetic systems.

Such geometrical responses of magnons without DM interactions were reported in a few studies. 
Scalar spin chirality is shown to support the thermal Hall effect for the honeycomb lattice by assuming a particular ground state spin configuration~\cite{Owerre2017TopologicalLattice}, 
and for the kagome lattice by incorporating a third neighbor coupling~\cite{GomezAlbarracin2021ChiralLattice,Owerre2017MagnonInteraction}. 
Another previous study reports that some organic materials~\cite{Naka2019SpinAntiferromagnets,Naka2020AnomalousAntiferromagnets,Naka2021PerovskiteGenerator} support geometrical magnon responses driven by temperature gradient due to special properties of dimers. Despite these previous studies, a guiding principle for realizing geometrical thermal responses of magnons without the DM interaction is still missing. In particular, the possibility of nontrivial magnon bands originating from an SU(2) gauge field without DM interaction has not been fully explored.

In this paper, we study geometrical thermal responses of magnons that do not rely on the DM interaction. Specifically, we focus on the antiferromagnetic Heisenberg model with  $AB$ sublattices. As the sublattice degrees of freedom enables to introduce the SU(2) gauge field to the magnons, this model is a suitable playground for pursuing the role of the non-Abelian gauge field on the geometrically nontrivial magnon bands. We first derive a general condition for generating nonzero Berry curvature without the DM interaction. We find that an effective $PT$-symmetry should be broken for obtaining geometrically nontrivial bands, and a noncollinear spin structure is necessary to break this $PT$-symmetry. From this viewpoint, 
frustrated spin systems are suitable for pursuing noncollinear spin configurations~\cite{2011IntroductionMagnetism,Zelezny2017Spin-PolarizedAntiferromagnets,Kimata2019MagneticAntiferromagnet}.
%In addition, the honeycomb lattice is the simplest lattice structure that host the $AB$ sublattices degrees of freedom.  Therefore, 
Thus we consider geometrical thermal responses in a frustrated honeycomb spin systems as a simple example. Specifically, we study the  $J_1$-$J_2$-$J_2^\prime$ Heisenberg model on the honeycomb lattice. The frustration naturally leads to the spiral order in the ground state and support the non-zero Berry curvature.
Furthermore, we also consider spin transport enabled by nontrivial magnon bands, i.e., the magnon spin Nernst effect. In the noncollinear system, the magnon spin Nernst effect is governed by a quantity that is closely related to the Berry curvature~\cite{Li2020IntrinsicAntiferromagnet,Park2020ThermalInteraction}. We demonstrate that the frustrated honeycomb Heisenberg model also supports the magnon spin Nernst effect without DM interaction.

The rest of this paper is organized as follows. In Sec.~II, we study magnon excitations using the Holstein-Primakoff transformation for the spiral phase with $AB$ sublattices, and derive the symmetry condition that the Berry curvature and the thermal Hall conductivity appear. In Sec.~III, we consider the $J_1$-$J_2$-$J_2^\prime$ model on the honeycomb lattice and study the thermal Hall conductivity. 
In Sec.~IV, we study the spin Nernst effect of $J_1$-$J_2$-$J_2^\prime$ model. 
In Sec.~V, we present a brief discussion.

\section{MAGNON HAMILTONIAN IN SPIRAL PHASE}
\label{magnon_sec}

In this section, we consider the condition for the non-zero Berry curvature and the thermal Hall conductivity in the $AB$ sublattice systems. First, in order to calculate the thermal Hall effect, we review the magnon expansion in $AB$ sublattice systems. Then, we introduce the formulation of the thermal Hall effect of magnons. After these preparations, we derive a general condition for generating non-zero thermal Hall effect.
\subsection{Magnon Hamiltonian of $AB$ sublattices}
We study the magnon Hamiltonian of the system with $AB$ sublattices. To obtain the magnon Hamiltonian, we perform the Holstein-Primakoff transformation for the spin $S$ systems~\cite{Holstein1940FieldFerromagnet},
\begin{equation} \label{HPtrans}
\begin{cases} 
    S^{\prime +}_i\simeq\sqrt{2S}a_i,
    S^{\prime -}_i\simeq\sqrt{2S}a^\dagger_i,
    S^{\prime z}_i=S-a^\dagger_i a_i
    &   \text{for $i \in A$}  \\
    S^{\prime +}_i\simeq\sqrt{2S}b_i,
    S^{\prime -}_i\simeq\sqrt{2S}b^\dagger_i,
    S^{\prime z}_i=S-b^\dagger_i b_i
    &   \text{for $i \in B$}
\end{cases},
\end{equation}    
where $a^\dagger_i$ and $b^\dagger_i$ are bosonic creation operators, $\bm{S}^\prime$ is a spin operator along the spin configuration of the ground state, and $S^{\prime\pm}_i=S_i^{\prime x}\pm iS_i^{\prime y}$. For a system in which the ground state is not ferromagnetic, the magnon Hamiltonian contains $\alpha^\dagger_i\alpha^\dagger_j$ and $\alpha_i\alpha_j$ terms with $\alpha_i$ being $a_i$ or $b_i$. Thus, after the Fourier transformation, 
we obtain the magnon Hamiltonian as a $4\times4$ matrix,
\begin{equation}
H=\frac{1}{2}\sum_{\bm{k}}\Psi^\dagger(\bm{k})H(\bm{k})\Psi(\bm{k}).
\end{equation}
This type of Hamiltonian is called the Bogoliubov–de Gennes (BdG) Hamiltonian. Here, $\Psi(\bm{k})$ and $H(\bm{k})$ are
\begin{align}
    \Psi(\bm{k})&=(a(\bm{k}),b(\bm{k}),a^\dagger(-\bm{k}),b^\dagger(-\bm{k}))^T, \label{def_psi}\\
H(\bm{k})&=
\begin{pmatrix}
          \Xi(\bm{k}) & \Pi(\bm{k}) \\
          \Pi^*(-\bm{k}) & \Xi^*(-\bm{k}) \\
        \end{pmatrix},\label{eq:BdG}
\end{align}
where $\Xi(\bm{k})$ and $\Pi(\bm{k})$ are $2\times2$ matrices that satisfy $\Xi^\dagger(\bm{k})=\Xi(\bm{k})$, $\Pi^\dagger(\bm{k})=\Pi^*(-\bm{k})$.
Using Pauli matrices, we can write $\Xi(\bm{k})$ and $\Pi(\bm{k})$ as 
\begin{align}
    \Xi(\bm{k})&=\Xi^0(\bm{k})\sigma_0+\Xi^x(\bm{k})\sigma_x+\Xi^y(\bm{k})\sigma_y+\Xi^z(\bm{k})\sigma_z, \label{xi_ham}\\
    \Pi(\bm{k})&=\Pi^0(\bm{k})\sigma_0+\Pi^x(\bm{k})\sigma_x+\Pi^y(\bm{k})\sigma_y+\Pi^z(\bm{k})\sigma_z, \label{delta_ham}
\end{align}
with $\Xi^i\in \mathbb{R}$ and $\Pi^i\in\mathbb{C}$ ($i=0,x,y,z$).
The BdG Hamiltonian should be diagonalized using a paraunitary matrix $T(\bm{k})$, which satisfies
\begin{align}
&T^\dagger(\bm{k})\sigma_3 T(\bm{k})=\sigma_3, \nonumber \\
    &\sigma_3=\begin{pmatrix}
              1 & 0 & 0 & 0 \\
              0 & 1 & 0 & 0 \\
              0 & 0 & -1 & 0 \\
              0 & 0 & 0 & -1 \\
    \end{pmatrix},
\end{align}
so as to retain the canonical commutation relation for the transformed magnon operator $T^{-1}(\bm{k})\Psi(\bm{k})$.
The eigenvalues have the following form due to the inherent particle-hole symmetry as 
\begin{align}
    T^\dagger(\bm{k})H(\bm{k})T(\bm{k})&=E(\bm{k})\notag
    \\&=\textrm{diag}(E_1(\bm{k}),E_2(\bm{k}),E_1(-\bm{k}),E_2(-\bm{k})). \label{diagonalize}
\end{align}
Applying $T(\bm{k})\sigma_3$ to Eq.~($\ref{diagonalize}$), we obtain
\begin{equation}
    \sigma_3H(\bm{k})T(\bm{k})= T(\bm{k})\sigma_3 E(\bm k) \label{eigen_T}.
\end{equation}
Namely, we can obtain $T(\bm{k})$ as eigenvectors of $\sigma_3 H$:
If we write the paraunitary matrix $T(\bm{k})$ as
\begin{equation}
    T(\bm{k})=(\bm{t}_1(\bm{k}),\bm{t}_2(\bm{k}),\bm{t}_3(\bm{k}),\bm{t}_4(\bm{k})),
\end{equation}
we can write Eq.~($\ref{eigen_T}$) in the form of an eigenvalue problem for $\sigma_3 H$ as
\begin{equation}
    \sigma_3H(\bm{k})\bm{t}_n(\bm{k})=(\sigma_3E(\bm{k}))_{nn}\bm{t}_n(\bm{k}).
\end{equation}

\subsection{Thermal Hall effect and Berry curvature}
We calculate the thermal Hall  conductivity by using the linear response theory. The temperature gradient is written as $T(\bm{r})=T_0(1-\chi(\bm{r}))$, where $T_0$ is a constant temperature and $\chi$ is a small parameter with a zero average. 
We write the thermal Hall current $J^Q_{\mu}$ as
\begin{equation*}
    J^Q_{\mu}=L_{\mu\nu}\left(T\nabla_\nu\frac{1}{T}-\nabla_\nu \chi \right),
\end{equation*} 
and the thermal Hall conductivity $\kappa_{\mu\nu}$ as
\begin{equation*}
    \kappa_{\mu\nu}=\frac{L_{\mu\nu}}{T}.
\end{equation*}

From a continuity equation, we can calculate the thermal current, and using the Kubo formula, we can write the thermal Hall conductivity $\kappa_{\mu\nu}$ as \cite{Matsumoto2014ThermalInteraction}
\begin{align}
    \kappa_{\mu\nu}&=-\frac{k_B^2 T}{\hbar}\sum_{n=1,2}\int_{BZ}\frac{dk^2}{(2\pi)^2}\Omega(\bm{k})_{n,\mu\nu} \nonumber \\
    &\hspace{7em} \times \left(c_2(\rho(E_n(\bm{k})))-\frac{\pi^2}{3}\right),
    \label{thermal_hall}
\end{align}
where
\begin{align*}
    c_2(x)&=\int_0^x dt \left(\log{\frac{1+t}{t}} \right)^2\\
    &=(1+x)\left(\log\frac{1+x}{x} \right)^2-(\log x)^2-2\textrm{Li}_2(-x),
\end{align*}
and $\textrm{Li}_n(x)$ is polylogarithm function. $\Omega(\bm{k})_{n,\mu\nu}$ is the Berry curvature of the $n$-th magnon band, 
\begin{equation}
\Omega_{n,\mu\nu}(\bm{k})=-2\textrm{Im}\left[\sigma_3\frac{\partial T^\dagger(\bm{k})}{\partial k_\alpha}\sigma_3\frac{\partial T(\bm{k})}{\partial k_\beta}\right]_{nn},\label{omega}
\end{equation}
which measures a nontrivial band geometry.

\subsection{Effective $PT$ and $T$ symmetries}
Symmetry plays an important role in the emergence of nontrivial magnon bands with Berry curvature.
In particular, we find that the Berry curvature of magnon bands vanishes under an effective $PT$ symmetry, in a similar manner to the Berry curvature in electronic systems.
In this subsection, we derive a symmetry condition for the non-zero Berry curvature and thermal Hall conductivity.

Let us suppose that the system has a symmetry given by
\begin{equation}
    P^{\dagger} H^*(\bm{k})P=H(\bm{k}) \label{symmetry}
\end{equation}
with a paraunitary matrix $P$ satisfying $P^\dagger\sigma_3P=\sigma_3$.
By utilizing Eq.~($\ref{symmetry}$), we can rewrite Eq.~($\ref{eigen_T}$) as
\begin{equation}
    \sigma_3H(k)P^*T^*(\bm{k})=P^*T^*(\bm{k})\sigma_3E(\bm{k}),
\end{equation}
which implies that $P^*T^*({\bm{k}})$ satisfies the same equation ($\ref{eigen_T}$) for $T(\bm{k})$. Namely, if there is no degeneracy, $T(\bm{k})$ should satisfy
\begin{equation}
T(\bm{k})=P^*T^*(\bm{k})M_{\bm{k}},  \label{T_symmetry} 
\end{equation}
where $(M_{\bm{k}})_{j,l}=\delta_{j,l}\exp[i\theta_{j,\bm{k}}]$ comes from the fact that we can choose the overall phases of the eigenvectors arbitrarily.
\par 

We investigate how this symmetry operation affects the Berry curvature. Considering the condition (\ref{T_symmetry}), the Berry curvature (\ref{omega}) can be written as
\begin{align}
\notag
    \Omega_{n,\alpha\beta}(\bm{k})&=-2\textrm{Im}\left[\sigma_3\frac{\partial T^\dagger(\bm{k})}{\partial k_\alpha}\sigma_3\frac{\partial T(\bm{k})}{\partial k_\beta}\right]_{nn}\\ \notag
&=-2\textrm{Im}\left[\sigma_3\frac{\partial M_{\bm{k}}^\dagger T^{\dagger *}(\bm{k})}{\partial k_\alpha}P^{*\dagger}\sigma_3 P^*\frac{\partial T^*(\bm{k})M_{\bm{k}}}{\partial k_\beta}\right]_{nn}\\
&=2\textrm{Im}\left[\sigma_3\frac{\partial T^{\dagger}(\bm{k})}{\partial k_\alpha}\sigma_3\frac{\partial T(\bm{k})}{\partial k_\beta}\right]_{nn}
\nonumber\\
&=-\Omega_{n,\alpha\beta}(\bm{k}).
\end{align}
Namely, the Berry curvature becomes zero under the symmetry ($\ref{symmetry}$). \par
Even if the Berry curvature takes nonzero value, the thermal Hall conductivity can vanish in some cases when the integral in Eq.~(\ref{thermal_hall}) has a cancellation.
Especially, when the Hamiltonian satisfy the effective time reversal symmetry (effective TRS) 
\begin{equation}
\tilde{P}^{\dagger} H^*(\bm{k})\tilde{P}=H(-\bm{k})    \label{effetive_TRS}
\end{equation}
with a paraunitary matrix $\tilde{P}$, the paraunitary matrix $T(\bm{k})$ obeys the condition $T(\bm{k})=\tilde{P}^*T^*(-\bm{k})M_{\bm{k}}$ and the Berry curvature $\Omega_{n,xy}(\bm{k})$ satisfies the relation $\Omega_{n,xy}(\bm{k})=-\Omega_{n,xy}(-\bm{k})$ \cite{Mook2019ThermalAntiferromagnets}. 
The effective TRS also imposes $E_n({\bm{k}})=E_n(-{\bm{k}})$ and $c_2(\rho(E_n(\bm{k})))=c_2(\rho(E_n(-\bm{k})))$. From these, the integrand of Eq.~(\ref{thermal_hall}) is odd in $\bm{k}$, and thus the thermal Hall conductivity $\kappa_{\alpha\beta}$ vanishes.\par

\subsection{Spiral phase}
To obtain nonzero thermal Hall conductivity, we need to break the effective $PT$ and $T$ symmetries. Here we consider $AB$-sublattice systems in the spiral phase, and discuss the general condition for breaking the symmetries and specific examples of symmetry-breaking interactions. To this end,
we consider the spin Hamiltonian 
\begin{equation}
\label{generalisedmodel}
    H=H_{J}+H_{\Delta}+H_h.
\end{equation}
Here, the first term
\begin{equation}
    H_{J}=\sum_{i\neq j}J_{{\alpha}{\beta}}(\bm{r})\bm{S}_i\cdot\bm{S}_j
\end{equation}
denotes the Heisenberg interaction between spins $\bm{S}_i$ and $\bm{S}_j$ with the coupling $J_{{\alpha}{\beta}}(\bm{r})$, where $\alpha,\beta=A,B$ represent the sublattice to which $i$ and $j$ sites belong, respectively, and $\bm{r}$ represents the distance between $i$ and $j$ sites.
The second term $H_{\Delta}$  
is the easy-axis anisotropy part,
\begin{align}
    H_{\Delta}&=\sum_i \Delta_\alpha (S^z_i)^2, \\
    &=\Delta_A\sum_{i\in A}(S^z_i)^2+\Delta_B\sum_{i\in B}(S^z_i)^2.
\end{align}
The last term $H_{h}$ is a Zeeman coupling term,
\begin{equation}
    H_{h}=h\sum_i S^z_i.
\end{equation}

We assume that the spin configuration of classical ground state is given by 
\begin{equation}
\bm{S}_{i}=S(\cos{\psi_i}\cos{(\bm{Q}\cdot\bm{R}_i+\phi_i)},\cos{\psi_i}\sin{(\bm{Q}\cdot\bm{R}_i+\phi_i)},\sin{\psi_i}) , \label{spin_a}
\end{equation}
where $\psi_i\in [-\pi/2,\pi/2]$ describes the canting angle from the $xy$ plane, and $\bm{Q}$ represents a pitch of the spiral. The canting angle $\psi_i$ is $\psi_A$ ($\psi_B$) if $i$ is in the $A$ ($B$) sublattice. Similarly, we assume that $\phi_i=\phi_{\alpha}$ for $i\in \alpha$ with $\alpha=A,B$. The position $\bm{R}_i$ denotes the center of the unit cell which contains the site $i$.

Because the spin Hamiltonian is symmetric with respect to the rotation of spin around the $z$-axis,
hereafter we set $\phi_A=0,$ $\phi_B=\phi$ without loss of generality, 
with which $\phi$ describes an in-plane angle between two spins in the same unit cell. This ansatz generally describes noncollinear spin configurations with single $\bm{Q}$. 

For the present canted spins, a new spin coordinate $\bm{S'}$ along the ground state spin configuration can be written as~\cite{Zhitomirsky1998MagnetizationAntiferromagnet, Owerre2017TopologicalLattice}
\begin{widetext}
\begin{align}
\bm{S}_{i}&=R^z(\bm{Q}\cdot\bm{R}_i+\phi_i)R^y(\pi/2-\psi_i)\bm{S}^\prime_{i} \notag\\
&=
\begin{pmatrix}
          \sin{\psi_i}\cos{(\bm{Q}\cdot\bm{R}_i+\phi_i)} & -\sin{(\bm{Q}\cdot\bm{R}_i+\phi_i)} & \cos{\psi_i}\cos{(\bm{Q}\cdot\bm{R}_i+\phi_i)} \\
            \sin{\psi_i}\sin{(\bm{Q}\cdot\bm{R}_i+\phi_i)} & \cos{(\bm{Q}\cdot\bm{R}_i+\phi_i)} & \cos{\psi_i}\sin{(\bm{Q}\cdot\bm{R}_i+\phi_i)}\\
            -\cos{\psi_i} & 0 & \sin{\psi_i} \\
\end{pmatrix}
\begin{pmatrix}
        {S}^{\prime x}_{i}\\
        {S}^{\prime y}_{i}\\
        {S}^{\prime z}_{i}
\end{pmatrix}
        , \label{rotation}
\end{align}
\end{widetext}
where $R^k(\theta)$ denotes a spin rotation operator with respect to the $k$-axis by $\theta$. 
Further rewriting $\bm{S}^\prime$ with the magnon operators using Holstein-Primakoff transformation (\ref{HPtrans})
and substituting it to the spin Hamiltonian ($\ref{generalisedmodel}$), we obtain the $4\times4$ BdG Hamiltonian (\ref{eq:BdG}) for the present system. For the detailed form of $\Xi(\bm{k})$ and $\Pi(\bm{k})$, see Appendix.

Let us discuss the presence/absence of the effective $PT$ symmetry for the present case. 
Here, for simplicity, we assume a lattice structure where the $A$ and $B$ sublattices are interchanged upon spatial inversion (e.g., honeycomb lattice).
First, we note that the physical $T$ and $PT$ symmetries are explicitly broken due to the Zeeman field term $H_h$.
However, this term is invariant under the combination of $T$ or $PT$ operation with $\pi$ rotation of spin around $y$-axis.
On the other hand, the ground state spin configuration typically has a lower symmetry than the Hamiltonian, and indeed the spiral spin order is not invariant under the above symmetry operation.
We here consider a symmetry operation $X$, which is obtained by further combining $\phi_A+\phi_B$ rotation of spin around the $z$-axis (to the $PT$ operation and $\pi$ rotation around $y$-axis). The spin configuration is transformed under $X$ as
\begin{align*}
    &S(\cos{\psi_i}\cos{(\bm{Q}\cdot\bm{R}_i+\phi_i)},\cos{\psi_i}\sin{(\bm{Q}\cdot\bm{R}_i+\phi_i)},\sin{\psi_i})\\
    &\rightarrow S(\cos{\psi_{-i}}\cos{(\bm{Q}\cdot\bm{R}_i+\phi_{i})},\cos{\psi_{-i}}\sin{(\bm{Q}\cdot\bm{R}_i+\phi_i)},\sin{\psi_{-i}}),
\end{align*}
where $-i \in B (A)$ if $i \in A (B)$. 
This implies that the ground state does not change under $X$ when $\psi_A=\psi_B$,
and the magnon Hamiltonian should have a corresponding symmetry if the spin Hamiltonian is also symmetric with respect to $X$.

Now, let us consider how this symmetry operation $X$ acts
on the magnon Hamiltonian. To this end, first we consider the transformation
for the spin operator, 
\begin{align*}
\bm{S}_{i}\to X\bm{S}_{i} & =R^{y}(\pi)R^{z}(-\phi_{A}-\phi_{B})(-\bm{S}_{-i})\\
 & =-R^{y}(\pi)R^{z}(-\bm{Q}\cdot\bm{R}_{i}-\phi_{i})R^{y}(\pi/2-\psi_{-i})\bm{S}_{-i}^{\prime}.
\end{align*}
On the other hand, the transformed spin operator can also be expressed
using the spin coordinate along the transformed ground state $X\bm{S}_{i}^{\prime}$
as 
\[
X\bm{S}_{i}=R^{z}(\bm{Q}\cdot\bm{R}_{i}+\phi_{i})R^{y}(\pi/2-\psi_{i})(X\bm{S}_{i}^{\prime}).
\]
Namely, when $\psi_{i}=\psi_{-i}$, $\bm{S}_{i}^{\prime}$ is transformed
as 
\[
X\bm{S}_{i}^{\prime}=\begin{pmatrix}1 & 0 & 0\\
0 & -1 & 0\\
0 & 0 & 1
\end{pmatrix}\bm{S}_{-i}^{\prime}
\]
 under $X$. Considering the fact that only the $y$ component
$S_{i}^{\prime y}=i\sqrt{S/2}(\alpha_{i}^{\dagger}-\alpha_{i})$ has
the imaginary coefficient to the magnon operators and that the sublattices are interchanged upon spatial inversion, we can express
the symmetry operation $X$ for the magnon Hamiltonian as
$H(\bm{k})\to P^\dagger H^\ast(\bm{k})P$ with \begin{equation}
    P=\begin{pmatrix}
             0 & 1 & 0 & 0\\
             1 & 0 & 0 & 0\\
             0 & 0 & 0 & 1\\
             0 & 0 & 1 & 0\\
    \end{pmatrix}.\label{specific_P}
\end{equation}

For the magnon Hamiltonian ($\ref{xi_ham},\ref{delta_ham}$), Eq.~($\ref{symmetry}$) is satisfied if
\begin{align}
\Xi^z=\Pi^z=0,\,\Pi^i\in\mathbb{R}.\label{eq:general-condition}
\end{align}

Let us discuss when the above condition can be broken, based on the detailed form of the magnon Hamiltonian given in Appendix.
For the Heisenberg interaction $H_J$, $\Xi^z$ and $\Pi^z$ are non-zero when $J_{AA}\neq J_{BB}$ (Eqs.~(\ref{xi_z}), (\ref{pi_z})). Furthermore, when $\psi_A\neq\psi_B$, $\textrm{Im}\Pi^x$ and $\textrm{Im}\Pi^y$ are also non-zero (Eqs.~(\ref{pi_x}), (\ref{pi_y})). For the anisotropy part $H_\Delta$, $\Pi^z$ is non-zero when $\Delta_{A}\neq \Delta_{B}$ (Eq.~(\ref{pi_z0})). Thus, when $A$ sites and $B$ sites are inequivalent, the Berry curvature can be non-zero. 

Furthermore, we consider the presence/absence of the effective TRS (\ref{effetive_TRS}), since breaking the effective TRS is necessary for non-zero thermal Hall conductivity. 
In particular, we focus on the simplest case of $\tilde{P}=I$ ($I$: an identity matrix) in the following.
We need $i\cos{k}$ or $\sin{k}$ terms to break the effective TRS (\ref{effetive_TRS}), and these terms of the BdG Hamiltonian for the spiral phase depend on $\sin{\bm{Q}\cdot{R}+\phi}$ or $\sin{\bm{Q}\cdot{R}}$. Thus, effective TRS is broken when the spin configuration satisfy $\sin{\bm{Q}\cdot{R}+\phi}\neq0$ or $\sin{\bm{Q}\cdot{R}}\neq0$. These conditions necessitate $\bm{Q}\cdot{R}\neq0,\pi$ or $\phi\neq0, \pi$. Hence, we need spiral configuration or nontrivial in-plane canting angle $\phi$ for non-zero thermal Hall conductivity besides the non-zero Berry curvature.

\subsection{SU(2) gauge fields in magnon bands}
\label{SU(2)sec}
In previous studies on thermal Hall responses  of magnetic systems \cite{Park2019TopologicalAntiferromagnets,Laurell2018MagnonInteractions,Doki2018SpinAntiferromagnet,Mook2019ThermalAntiferromagnets,Owerre2017TopologicalAntiferromagnets,Owerre2017NoncollinearInsulator}, the DM interaction is incorporated to generate nonzero Berry curvature of magnon bands. In this subsection, we comment on the role of the DM interaction in view of the symmetry condition [Eq.~(\ref{symmetry})] and the effective SU(2) gauge field. Specifically, we show that the presence of the DM interaction can break the symmetry  (\ref{symmetry}), and discuss how the similar SU(2) gauge field is obtained without the DM interaction in the spiral phase with the sublattice inequivalence.

First, we consider the out-of-plane DM interaction
\begin{equation}
    H_{\text{DM}}=\sum_{i,j}D_{\alpha\beta}(\bm{S}_i\times\bm{S}_j)_z.
\end{equation}
For this DM interaction, when $D_{AA}\neq D_{BB}$, we can obtain non-zero $\Xi^z$ even if $\psi_A=\psi_B$ (see Appendix). Thus, DM interaction can break the symmetry (\ref{symmetry}) and generate the non-zero Berry curvature \cite{Owerre2017NoncollinearInsulator}. In addition, $\Xi^z$ can be non-zero even if the spin configuration is the collinear and $\psi_i$ is $\pm{\pi/2}$. In this case, the DM interaction can be taken into the Heisenberg coupling with a phase factor $\chi=\arctan{(D/J)}$
\begin{equation}
    J\bm{S}_i\cdot\bm{S}_j+D\bm{S}_i\times\bm{S}_j=J_\text{eff}(e^{i\chi} S^+_iS^-_j+e^{-i\chi} S^-_iS^+_j), \label{dm_chi}
\end{equation}
where $J_\text{eff}$ is an effective Heisenberg coupling $J_\text{eff}=\sqrt{J^2+D^2}$. Thus, the DM interaction adds the phase factor $\chi$ to the hopping and acts as the virtual magnetic field and induce the non-zero $\Xi^z$.

On the other hand, in-plane DM interaction can also induce a non-zero Berry curvature with a different mechanism.
The in-plane DM interaction can induce SU(2) gauge field in canted spin systems, which is a non-Abelian gauge field with respect to the sublattice degrees of freedom in the magnon representation \cite{Kawano2019ThermalAntiferromagnet}. Now, we show that we can induce the SU(2) gauge field even without the DM interaction in a system with $\psi_A\neq\psi_B$. For simplicity, we consider the antiferromagnetic Heisenberg chain with nearest-neighbor coupling
\begin{equation*}
    H=\sum_{i\in A} (J\bm{S}_i\cdot\bm{S}_{i+1}+J^\prime\bm{S}_i\cdot\bm{S}_{i-1}).
\end{equation*}
Here we assume the spiral spin configuration given by (\ref{spin_a}). Again, we can set $\phi_A=0$ 
without loss of generality, and we obtain
\begin{align}
\bm{S}_i&=S(\cos{\psi_A}\cos{(\bm{Q}\cdot\bm{R}_i)},\sin{\psi_A}\sin{(\bm{Q}\cdot\bm{R}_i)},\sin{\psi_A}),\\ 
\bm{S}_{i+1}&=S(\cos{\psi_B}\cos{(\bm{Q}\cdot\bm{R}_i+\phi)},\sin{\psi_B}\sin{(\bm{Q}\cdot\bm{R}_i+\phi)},\sin{\psi_B})
\end{align}
for $i\in A$.

The spin Hamiltonian in the $\bm{S}^\prime$ coordinate can be obtained with Eq.~(\ref{rotation}) as follows,
\begin{align*}
        H=\sum_{i\in A}&[J_X S^{\prime x}_iS^{\prime x}_{i+1} +J_Y S^{\prime y}_iS^{\prime y}_{i+1} +J_Z S^{\prime z}_iS^{\prime z}_{i+1}\\ 
        &+D_0(S^{\prime x}_i S^{\prime y}_{i+1}-S^{\prime y}_i S^{\prime x}_{i+1})+D_1(S^{\prime x}_iS^{\prime y}_{i+1}+S^{\prime y}_iS^{\prime x}_{i+1})\\
        &+J^\prime_X S^{\prime x}_{i-1}S^{\prime x}_i +J^\prime_Y S^{\prime y}_{i-1}S^{\prime y}_i +J^\prime_Z S^{\prime z}_{i-1}S^{\prime z}_i\\ 
        &+D^\prime_0(S^{\prime x}_{i-1} S^{\prime y}_i-S^{\prime y}_{i-1} S^{\prime x}_i)+D^\prime_1(S^{\prime x}_{i-1}S^{\prime y}_i+S^{\prime y}_{i-1}S^{\prime x}_i)],
\end{align*}
where
\begin{align*}
    &J_X=J(\sin{\psi_A}\sin{\psi_B}\cos{\phi}+\cos{\psi_A}\cos{\psi_B}),\\
    &J_Y=J\cos{\phi},\\
    &J_Z=J(\cos{\psi_A}\cos{\psi_B}\cos{\phi}+\sin{\psi_A}\sin{\psi_B}),\\
    &D_0=-J\frac{\sin{\psi_A}+\sin{\psi_B}}{2}\sin{\phi},\\
    &D_1=J\frac{\sin{\psi_B}-\sin{\psi_A}}{2}\sin{\phi},\\
    &J^\prime_X=J^\prime(\sin{\psi_A}\sin{\psi_B}\cos{(\phi-\bm{Q}\cdot\bm{R})}+\cos{\psi_A}\cos{\psi_B}),\\
    &J^\prime_Y=J^\prime\cos{(\phi-\bm{Q}\cdot\bm{R})},\\
    &J^\prime_Z=J^\prime(\cos{\psi_A}\cos{\psi_B}\cos{(\phi-\bm{Q}\cdot\bm{R})}+\sin{\psi_A}\sin{\psi_B}),\\
    &D^\prime_0=J^\prime\frac{\sin{\psi_A}+\sin{\psi_B}}{2}\sin{(\phi-\bm{Q}\cdot\bm{R})},\\
    &D^\prime_1=J^\prime\frac{\sin{\psi_B}-\sin{\psi_A}}{2}\sin{(\phi-\bm{Q}\cdot\bm{R})}\\
\end{align*}
with $\bm{R}=\bm{R}_{i+1}-\bm{R}_i$. In these terms, $D_1$ and $D_1^\prime$ are non-zero only when the spin configuration satisfies $\psi_A\neq\psi_B$, which implies that these are the candidates for (effective) $PT$ breaking term. Using the HP transformation (\ref{HPtrans}), we can write the $D_1$ and $D^\prime_1$ terms in terms of the magnon operators as
\begin{align*}
    &\sum_{i\in A}\frac{i}{2}[D_1(a^\dagger_ib^\dagger_{i+1}-a_ib_{i+1}) +D^\prime_1(a^\dagger_ib^\dagger_{i-1}-a_ib_{i-1})] \\
    &=\sum_{i\in A}-\frac{1}{2}\left[D_1(a^\dagger_i,b_{i+1})\sigma_y
    \begin{pmatrix}
             a_i\\
             b^\dagger_{i+1}
    \end{pmatrix} 
    +D^\prime_1(a^\dagger_i,b_{i-1})\sigma_y
    \begin{pmatrix}
             a_i\\
             b^\dagger_{i-1}
    \end{pmatrix}\right].
\end{align*}
Then, after the Fourier transformation, we obtain 
\begin{align}
    \sum_k&\left[-\frac{D_1+D^\prime_1}{4}\cos{k}(a^\dagger_k,b_{-k})\sigma_y
    \begin{pmatrix}
             a_k\\
             b^\dagger_{-k}
    \end{pmatrix}\right.\notag\\
    &\left.
    -i\frac{D_1-D^\prime_1}{4}\sin{k}(a^\dagger_k,b_{-k})\sigma_y
    \begin{pmatrix}
             a_k\\
             b^\dagger_{-k}
    \end{pmatrix} \right].
\end{align}
Here, the second term contains $i\sigma_y$ and this is the origin of the non-zero $\Im{\Pi_y}$ for the BdG Hamiltonian. 
This term is the same form as a Rashba spin-orbit term, since we can see $(a^\dagger_k,b_{-k})$ as a pseudospinor operator \cite{Kawano2019ThermalAntiferromagnet}. This Rashba-like term contains $D_1-D^\prime_1$, supporting non-zero SU(2) gauge field when $D_1\neq D^\prime_1$. The condition $D_1\neq D^\prime_1$ is satisfied when $\bm{Q}\cdot\bm{R}\neq 0,\pi$ or $J\neq J^\prime$ and $\phi\neq 0,\pi$. Hence, we need spiral configuration or non-zero canting angle $\phi$ and asymmetric bonds other than the condition $\psi_A\neq\psi_B$ for SU(2) gauge field. 

In previous studies, the thermal Hall effect without DM interactions is reported in a kagome lattice system~\cite{GomezAlbarracin2021ChiralLattice,Owerre2017MagnonInteraction} and a honeycomb lattice system with $\Im{\Pi^i}=0$~\cite{Owerre2017TopologicalLattice}. From the viewpoint of the above discussion, the non-zero Berry curvature in these previous studies is derived from the phase factor of hoppings as in the case of the out-of-plane DM interaction. 
In contrast, the SU(2) gauge field also induces non-zero Berry curvature as we have clarified above and demonstrate for the $J_1$-$J_2$-$J'_2$ models in the following.

\begin{figure}[tb]
\centering
\includegraphics[width=\linewidth]{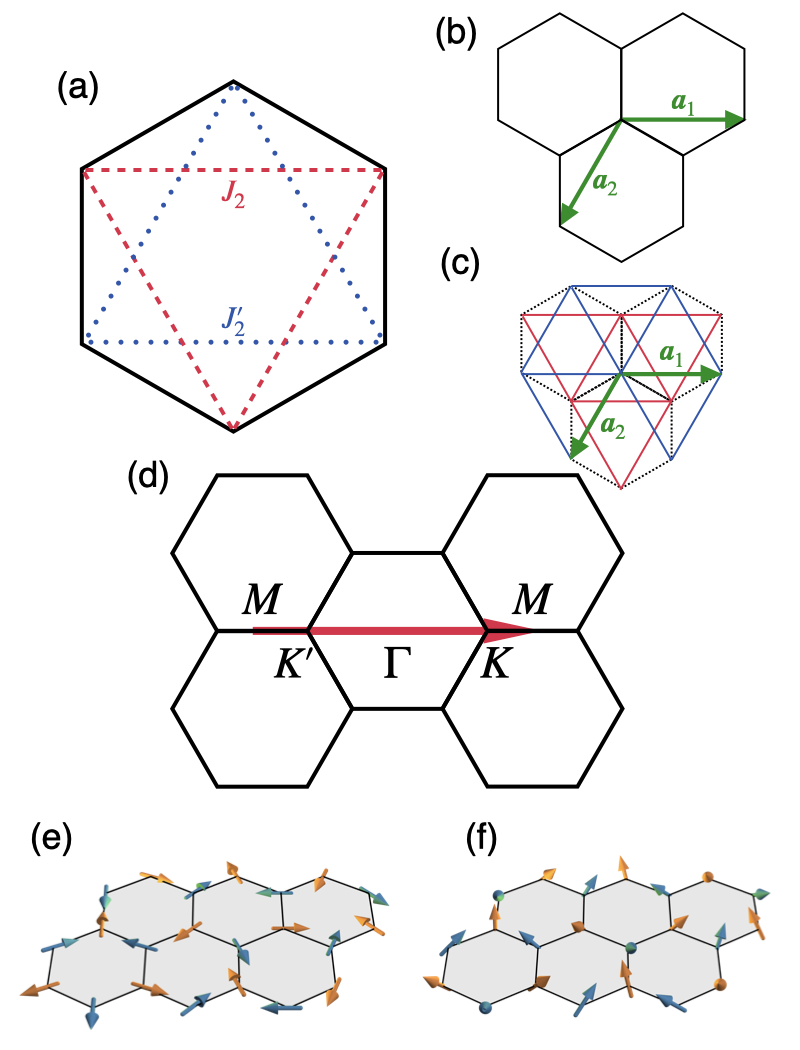}
\caption{(a) The $J_1$-$J_2$-$J_2^\prime$ model on the honeycomb lattice. (b) Vectors $\bm{a}_1$ and $\bm{a}_2$ of the honeycomb lattice. (c) Vectors $\bm{a}_1$ and $\bm{a}_2$ of bilayer triangle lattices. The red lines and blue lines each represent the top layer and bottom layers' triangle lattices.  (d) The reciprocal space of the $J_1$-$J_2$-$J_2^\prime$ model on the honeycomb lattice. (e) and (f) is the spin configuration of the model$\rm(\hspace{.08em}ii\hspace{.08em})$ with $J_1=1.0$, $J_2=2.0$, $J_2^\prime=2.4$, $\Delta_A=\Delta_B=0.05$. (e) $h=0$ and the canting angle $\psi_A=\psi_B=0$. (f) $h=8$ and the canting angle $\psi_A\neq\psi_B$.}
\label{fig:model}
\end{figure}

\begin{figure*}[htb]
\includegraphics[width=\linewidth]{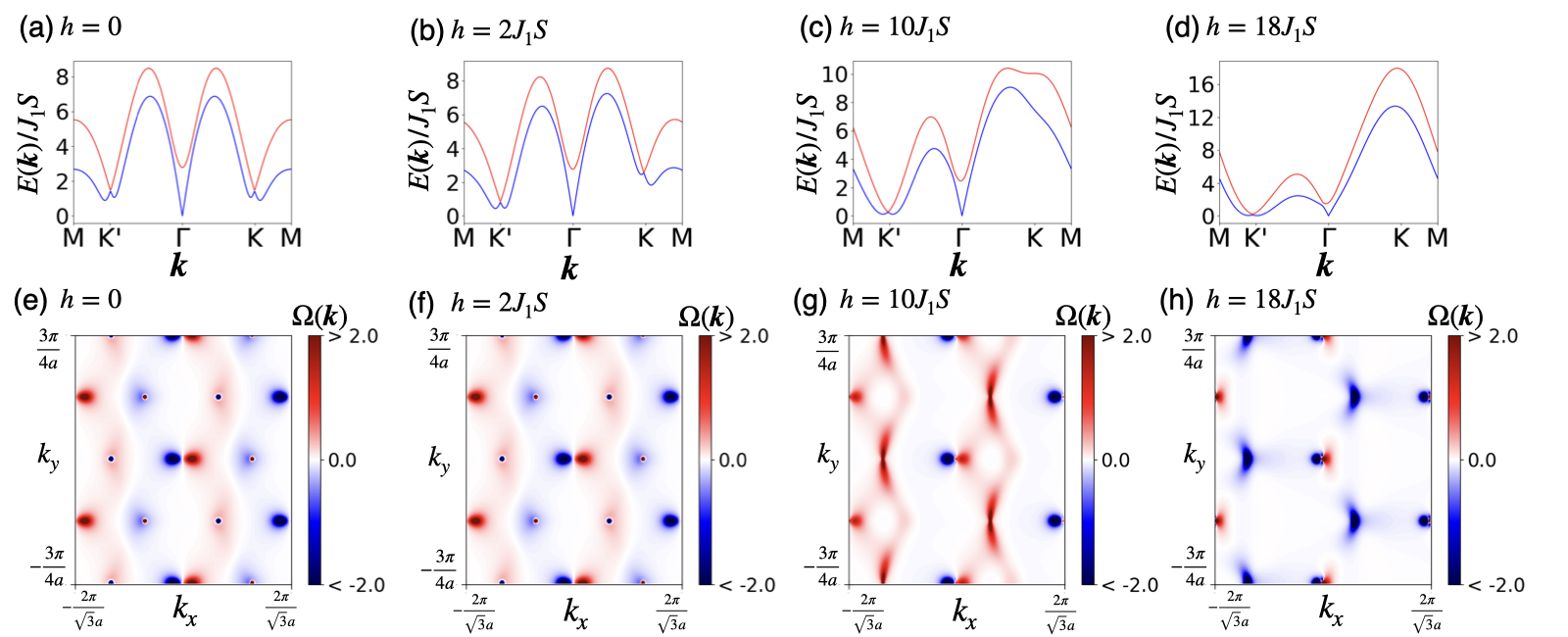}
\caption{\label{fig:J1J2J3_berry} The energy band and the Berry curvature $\Omega_{xy}$ of the model$\rm(\hspace{.18em}i\hspace{.18em})$ with $J_1=1.0$, $J_2=2.0$, $J_2^\prime=2.4$, $\Delta_A=\Delta_B=0.05$. (a, b, c, d) The energy bands for (a) $h=0$, (b) $h=2J_1S$, (c) $h=10J_1S$, and (d) $h=18J_1S$. (e,f,g,h) The Berry curvature $\Omega_{xy}$ of the lower band for (e) $h=0$, (f) $h=2J_1S$, (g) $h=10J_1S$, and (h) $h=18J_1S$. }
\end{figure*}

\section{MAGNON HAMILTONIAN IN $J_1$-$J_2$-$J_2^\prime$ MODEL}
\label{J1J2J3_sec}

Now, we demonstrate the thermal Hall effect without DM interaction. As we have clarified in the above section, we need inequivalent $AB$ sublattices, and $\bm{Q}\cdot{R}\neq 0$, $\pi$ or $\phi\neq 0,\pi$ for the thermal Hall effect. 
$J_1$-$J_2$-$J_2^\prime$ model on the honeycomb lattice is a simple example that satisfies above conditions. 
In this model, the next-nearest-neighbor coupling $J_2$ and $J_2^\prime$ induce frustration, which leads to a spiral order with $\bm{Q}\neq 0$ on the grand state spin configuration. 
In order to make the $A$ and $B$ sublattices inequivalent, we introduce the inequivalent next-nearest-neighbor coupling $J_2\neq J_2^\prime$ or inequivalent anisotropy $\Delta_A\neq\Delta_B$.

\subsection{Spin Hamiltonian}

We consider the Heisenberg model on the honeycomb lattice depicted in Fig.~\ref{fig:model}(a).
While the honeycomb lattice has $AB$ sublattices, we assume that these sublattices are inequivalent (e.g., composed of two different atoms), so that the coupling constants may take different values for $A$ and $B$ sublattices. Namely, here we consider the $J_1$-$J_2$-$J_2^\prime$ Heisenberg model, 
whose Hamiltonian is given by
\begin{equation}
    H=H_{J}+H_{\Delta}+H_h \label{model_ham},
\end{equation}
with
\begin{align}
    H_J&=J_{1}\sum_{\langle i,j \rangle}\bm{S}_i\cdot\bm{S}_j+J_{2}\sum_{\langle\langle i,j \rangle\rangle\in A}\bm{S}_i\cdot\bm{S}_j
    \nonumber\\
    &+J_{2}^\prime\sum_{\langle\langle i,j \rangle\rangle\in B}\bm{S}_i\cdot\bm{S}_j,
\end{align}
\begin{equation}
    H_{\Delta}=\Delta_A\sum_{i \in A}(S_i^z)^2+\Delta_B\sum_{i \in B}(S_i^z)^2,
\end{equation}
and
\begin{equation}
    H_h=-h\sum_iS_i^z.
\end{equation}
Here the index $i$ runs over all sites, and $\sum_{\langle i,j \rangle}$ and $\sum_{\langle\langle i,j \rangle\rangle}$ means that sum over nearest-neighbor and next-nearest-neighbor of the honeycomb lattice, respectively.
The operator $\bm{S}_i$ is a spin at site $i$, and $A$ and $B$ are sublattices of honeycomb lattice. Figure~\ref{fig:model}(b) shows the primitive lattice vectors $\bm{a}_1=(\sqrt{3}a,0)$ and $\bm{a}_2=(-\sqrt{3}a/2,-3a/4)$ with the lattice constant $a$ (Hereafter, we set $a=1$ for simplicity). Namely, $A$ sites are located at $\bm{r}=m\bm{a}_1+n\bm{a}_2$, while $B$ sites are located at $\bm{r}=m\bm{a}_1+n\bm{a}_2+(0,a)$. 
Figure~\ref{fig:model}(d) shows the reciprocal space of the $J_1$-$J_2$-$J_2^\prime$ model on the honeycomb lattice.

We note that this model can also be regarded as a bilayer triangular lattice system, by considering $A$ ($B$) sites as the top (bottom) layer [See Fig.~\ref{fig:model}(c)].
In this case, $\sum_{\langle i,j \rangle}$ and $\sum_{\langle\langle i,j \rangle\rangle}$ indicate sums over nearest-neighbor interlayer and intralayer couplings, respectively.
In particular, we emphasize that it is not necessarily unrealistic to consider a situation where $J_2$ and $J_2^\prime$ are much larger than $J_1$.

In the case of $h=0$ and $J_2=J_2^\prime$, the classical limit of this model is studied. If $J_2/J_1>1/6$, the ground state spin configuration is given as~\cite{Rastelli1979Non-simpleHamiltonians,Fouet2001AnLattice,Mulder2010SpiralLattice,Bishop2012TheModel,Bishop2015FrustratedParameters}
\begin{align}
\label{classical_gs1}
\bm{S}_i&=S(\cos{(\bm{Q}\cdot\bm{R}_i)},\sin{(\bm{Q}\cdot\bm{R}_i)},0) ~~~\textrm{for}~i\in A, \\
\label{classical_gs2} 
\bm{S}_{i}&=S(\cos{(\bm{Q}\cdot\bm{R}_i+\phi)},\sin{(\bm{Q}\cdot\bm{R}_i+\phi)},0)~~~\textrm{for}~i\in B.
\end{align} 
In the spiral phase, we can minimize the classical energy by taking
\begin{align}
    \bm{Q}&=\left(\frac{2}{\sqrt{3}a}\cos^{-1}{\left[\frac{J_1-2J_2}{4J_2}\right]},0,0\right),\\ 
    \phi&=\pi .
\end{align}
We note here that there are two other ground states rotated by $\pm\frac{2\pi}{3}$ in the honeycomb plane.\par

In the case of $h\neq0$ and $J_2\neq J_2^\prime$, we assume that classical ground states can be written as Eq.~($\ref{spin_a}$). 
Even for $J_2\neq J_2^\prime$, we assume $\phi_A=0$ and $\phi_B=\pi$, which is the known result for the $J_2=J_2^\prime$ case \cite{Mulder2010SpiralLattice}.
Namely, we write the classical ground states as
\begin{align}
\bm{S}_i&=S(\cos{\psi_A}\cos{(\bm{Q}\cdot\bm{R}_i)},\sin{\psi_A}\sin{(\bm{Q}\cdot\bm{R}_i)},\sin{\psi_A}) 
\nonumber\\
& \hspace{16em}~~~\textrm{for}~i\in A, 
\label{assume_gsa}\\ 
\bm{S}_i&=S(-\cos{\psi_B}\cos{(\bm{Q}\cdot\bm{R}_i)},-\sin{\psi_B}\sin{(\bm{Q}\cdot\bm{R}_i)},\sin{\psi_B}) \nonumber\\
& \hspace{16em}~~~\textrm{for}~i\in B. \label{assume_gsb}
\end{align}
Here, $\psi_A$ and $\psi_B$ are canting angles from the $xy$-plane, and we estimate $\bm{Q}$, $\psi_A$, and $\psi_B$ by minimizing the classical energy 
\begin{align}
E=&NS^2[-J_1\cos{\psi_A}\cos{\psi_B}(1+\cos{(Q_1+Q_2)}+\cos{Q_2}) \notag\\
&+(J_2\cos^2{\psi_A}+J_2^\prime\cos^2{\psi_B})\notag\\
&\times(\cos{Q_1}+\cos{Q_2}+\cos{(Q_1+Q_2)})\notag\\
&+3J_1\sin{\psi_A}\sin{\psi_B}+3(J_2\sin^2{\psi_A}+J_2^\prime\sin^2{\psi_B})\notag\\
&+\Delta_A\sin^2{\psi_A}+\Delta_B\sin^2{\psi_B}]
\notag\\
&-NS(h\sin{\psi_A}+h\sin{\psi_B}),
\end{align}
where $N$ is the site number of $A$ sites and $B$ sites, and $Q_1=\sqrt{3}aQ_x$, $Q_2=-\sqrt{3}a/2Q_x-3a/4Q_y$. For the case of $h=0$, due to the (easy-plane) magnetic anisotropy, the canting angles $\psi_A$ and $\psi_B$ are zero. Figure~\ref{fig:model}(e) shows the spin configuration of this case. When $h>0$, on the other hand, spins are canted from the $xy$-plane as shown in Fig.~\ref{fig:model}(f).

\begin{figure}[tb]
\includegraphics[width=\linewidth]{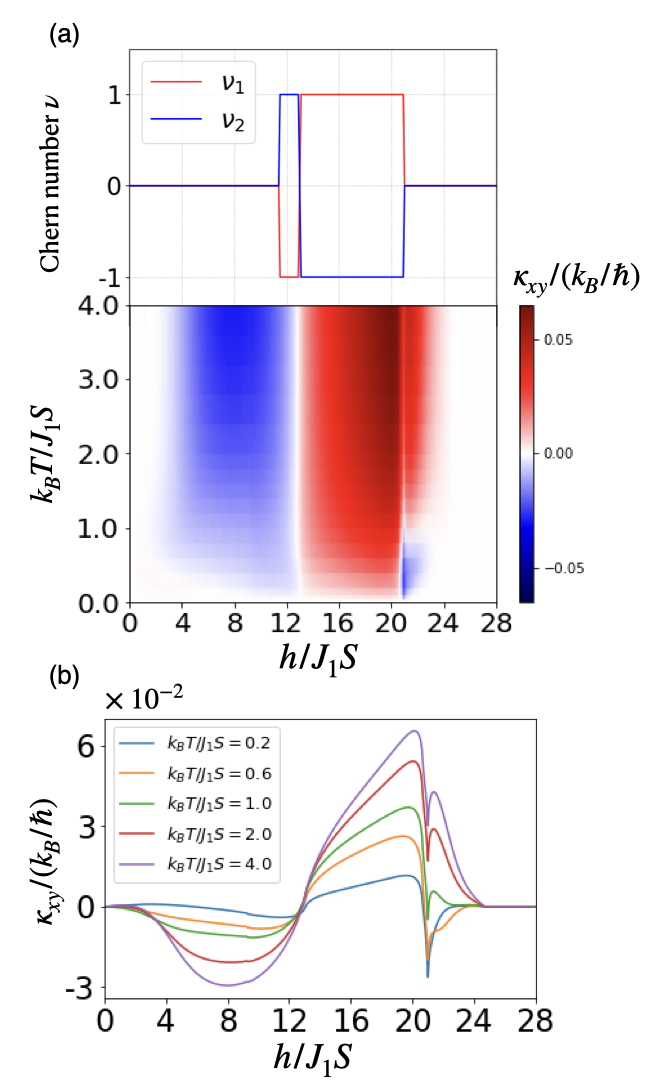}
\caption{\label{fig:J1J2J3_thermal} The Chern number $\nu$ and the thermal Hall conductivity $\kappa_{xy}$ of the model$\rm(\hspace{.18em}i\hspace{.18em})$ with $J_1=1.0$, $J_2=2.0$, $J_2^\prime=2.4$, $\Delta_A=\Delta_B=0.05$. Here, we calculate $\kappa_{xy}$ with $S=\frac{1}{2}$. 
(a) The magnetic field dependence of $\nu$ for each magnon band and the color plot of $\kappa_{xy}$. $\nu_1$ is the Chern number of the upper band, and $\nu_2$ is the Chern number of the lower band. 
The lower panel is the color plot of $\kappa_{xy}$. The sign of $\kappa_{xy}$ almost coincides with the sign of $\nu_1$. In particular, $\kappa_{xy}$ becomes zero and shows a sign change around  $h\sim13J_1S$, where the sign of the $\nu_1$ changes. 
(b) $\kappa_{xy}$ plotted as a function of the magnetic field for several temperatures.}
\end{figure}

\subsection{Magnon band, Berry curvature, and Chern number}

\begin{figure*}[!tb]
\includegraphics[width=0.9\linewidth]{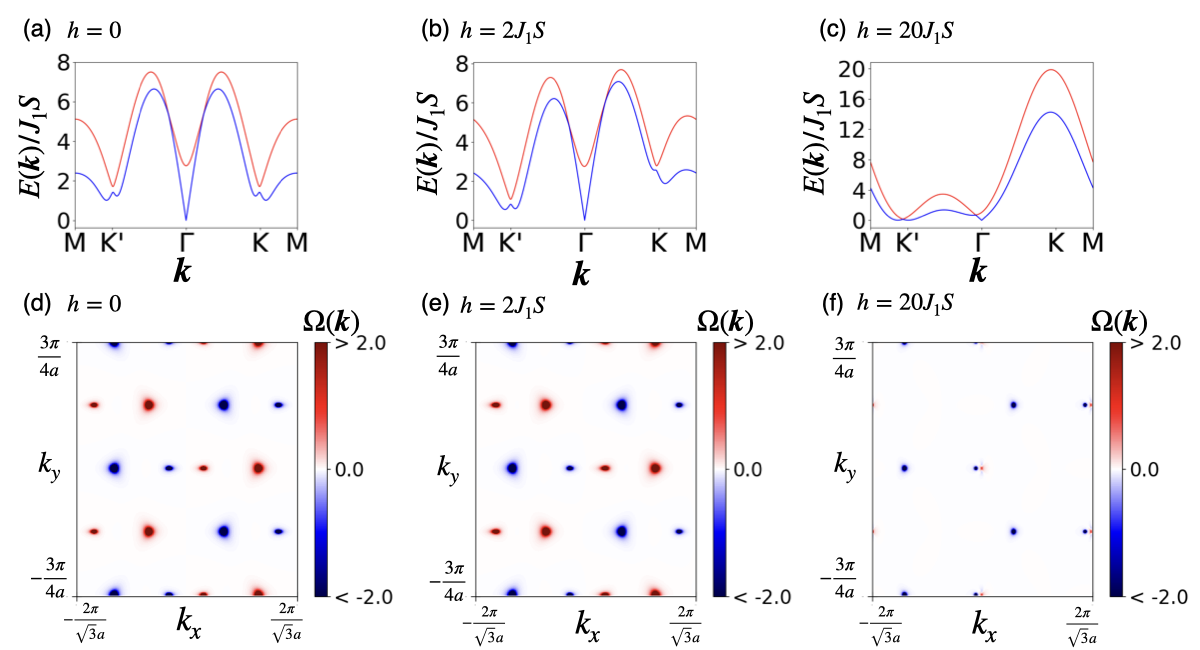}
\caption{\label{fig:J1J2_berry} 
The energy band abd the Berry curvature $\Omega_{xy}$ of the model$\rm(\hspace{.08em}ii\hspace{.08em})$ with $J_1=1.0$, $J_2=J_2^\prime=2.0$, $\Delta_A=0.05$, $\Delta_B=0.1$. (a), (b), and (c) is the energy band (a) at $h=0$, (b) at $h=2J_1S$, and (c) at $h=20J_1S$. (d), (e), and (f) show the Berry curvature of the lower band (d) at $h=0$, (e) at $h=2J_1S$, and (f) at $h=20J_1S$. The Berry curvature $\Omega_{xy}$ is large where the energy gap is small.}
\end{figure*}

\begin{figure}[!tb]
\includegraphics[width=0.9\linewidth]{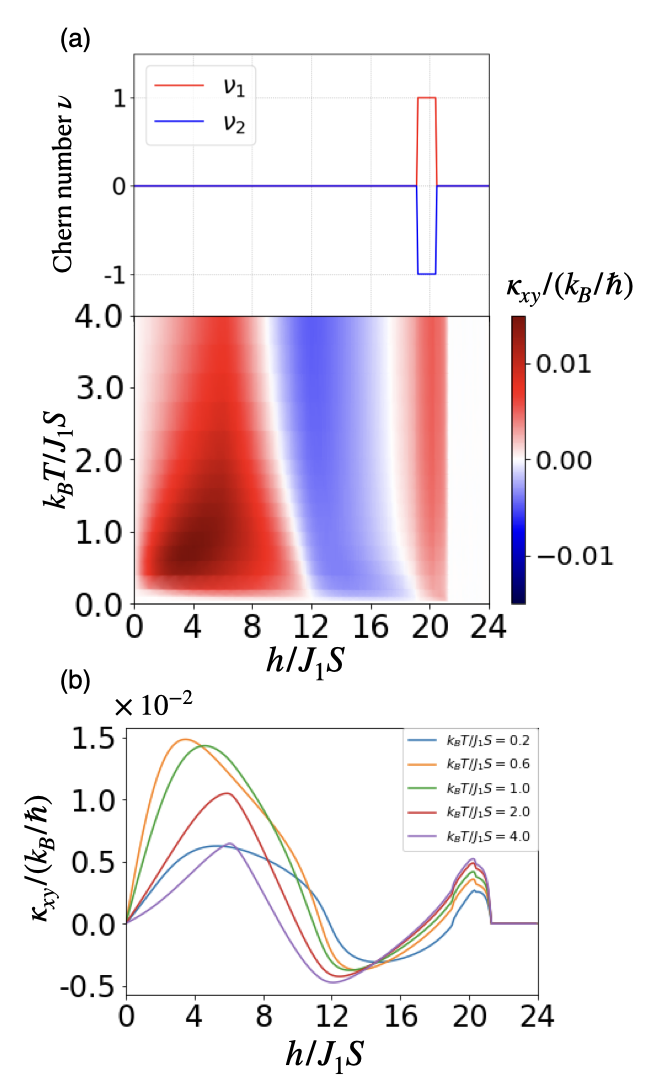}
\caption{\label{fig:J1J2_thermal} The Chern number $\nu$ and the thermal Hall conductivity $\kappa_{xy}$ of the model$\rm(\hspace{.08em}ii\hspace{.08em})$ with $J_1=1.0$, $J_2=J_2^\prime=2.0$, $\Delta_A=0.05$, $\Delta_B=0.1$. Here, we calculate the $\kappa_{xy}$ with $S=\frac{1}{2}$. 
(a) The magnetic field dependence of $\nu$ for each magnon band and the color plot of the thermal Hall conductivity. $\nu_1$ is the Chern number of the upper band, and $\nu_2$ is the Chern number of the lower band. $\kappa_{xy}$ becomes zero and shows a sign change around $h\sim18J_1S$, where the Chern number $\nu$ changes. 
(b) The thermal Hall conductivity plotted as the function of the magnetic field for several temperatures.}
\end{figure}

We apply Holstein-Primakoff transformation for the spin Hamiltonian ($\ref{model_ham}$) and obtain the magnon Hamiltonian (for details, see Appendix). 
In this model, there are two types of inequivalence introduced by $\rm(\hspace{.18em}i\hspace{.18em})$ inequivalent Heisenberg coupling for two triangular lattices ($J_2\neq J_2^\prime$) and $\rm(\hspace{.08em}ii\hspace{.08em})$ inequivalent anisotropy for two triangular lattices ($\Delta_A\neq\Delta_B$). Thus we name the model$\rm(\hspace{.18em}i\hspace{.18em})$, $J_2\neq J_2^\prime$, $\Delta_A=\Delta_B$, and the model$\rm(\hspace{.08em}ii\hspace{.08em})$, $J_2=J_2^\prime$, $\Delta_A\neq \Delta_B$. 
In the following, we first discuss the results for model$\rm(\hspace{.18em}i\hspace{.18em})$  
with $J_2\neq J_2^\prime$, 
and then proceed to the results for model$\rm(\hspace{.08em}ii\hspace{.08em})$ with $\Delta_A\neq \Delta_B$.

In Fig.~\ref{fig:J1J2J3_berry}, we show the energy band and the Berry curvature of model$\rm(\hspace{.18em}i\hspace{.18em})$, which has inequivalent Heisenberg coupling for two triangular lattices ($J_2\neq J_2^\prime$). Here, the energy band is plotted along the paths shown in Fig.~\ref{fig:model}(d). 

Figures.~\ref{fig:J1J2J3_berry}(a) and (e) show the energy band and the Berry curvature in the absence of the external magnetic field, $h=0$. From Fig.~\ref{fig:J1J2J3_berry}(a), we can see that the small gaps around $K$ and $K^\prime$ are energetically equivalent to each other.
In this case, band gaps open, and the Berry curvature is non-zero as shown in Fig.~\ref{fig:J1J2J3_berry}(e), although the thermal Hall conductivity vanishes because the magnon Hamiltonian satisfies the effective TRS $H^*(\bm{k})=H(\bm{-k})$.

We show the energy band of model$\rm(\hspace{.18em}i\hspace{.18em})$ ($J_2\neq J_2^\prime$) with magnetic field $h=2J_1S$, $h=10J_1S$, and $h=18J_1S$ in Figs.~\ref{fig:J1J2J3_berry}(b), (c), and (d), respectively. If we turn on the magnetic field $h\neq0$, the two small gaps around $K$ and $K^\prime$ points become energetically inequivalent, since the effective TRS is now broken [$H(\bm{k})\neq H^*(\bm{-k})$]. Figures~\ref{fig:J1J2J3_berry}(b), (c), and (d)  show that the energy around $K^\prime$ decreases when $h\neq0$, and energy around $K$ increases. These changes in the band structure produce changes in the Berry curvature. 

Figures~\ref{fig:J1J2J3_berry}(f), (g), and (h) show the Berry curvature with $h=2J_1S$, $h=10J_1S$, and $h=18J_1S$. In the $h=2J_1S$ case, the Berry curvature satisfies $\Omega_{n,xy}(\bm{k})\simeq-\Omega_{n,xy}(-\bm{k})$ similarly to the $h=0$ case, as one can see from Fig.~\ref{fig:J1J2J3_berry}(f). As shown in Figs.~\ref{fig:J1J2J3_berry}(c) and (d), as the magnetic field increases, the gap around $K$ becomes larger and, accordingly, the Berry curvature around $K$ becomes smaller as shown in Figs.~\ref{fig:J1J2J3_berry}(g) and (h). 

From magnetic field dependence of Berry curvatures, we can predict that the Chern number is zero when the magnetic field is small, while the nonzero Chern number is realized for larger $h$. 
We show the magnetic field dependence of the Chern number and the thermal Hall conductivity in Fig.~\ref{fig:J1J2J3_thermal}(a). From the upper figure of Fig.~\ref{fig:J1J2J3_thermal}(a), we see that the Chern number is non-zero when the magnetic field $h$ is large. The lower panel of Fig.~\ref{fig:J1J2J3_thermal}(a) shows the color plot of the thermal Hall conductivity, while Fig.~\ref{fig:J1J2J3_thermal}(b) shows the thermal Hall conductivity at several temperatures. While the thermal Hall conductivity is related to the Berry curvature via Eq.~($\ref{thermal_hall}$), unlike the Hall effect of electron systems, the thermal Hall effect of magnons is not quantized, because the function $c_2(\rho(E)$ in Eq.~($\ref{thermal_hall}$) is not the function like a step function. 
Nonetheless, the thermal Hall conductivity shows a behavior related to that of the Chern number. To see this, first we remark that the Berry curvature of the upper band $\Omega_{1,\alpha\beta}$, and the lower band $\Omega_{2,\alpha\beta}$ satisfy $\Omega_{1,\alpha\beta}\sim-\Omega_{2,\alpha\beta}$, and that  $-(c_2(\rho(E))-\frac{\pi^2}{3})$ in Eq.~($\ref{thermal_hall}$) is a monotonously increasing function. These imply that the sign of the thermal Hall conductivity corresponds to the sign of the Chern number of the upper band. Especially, Figs.~\ref{fig:J1J2J3_thermal}(a) and (b) show that the sign of the thermal Hall conductivity changes reflecting the sign change of the Chern number.

The thermal Hall effect also appears in the model$\rm(\hspace{.08em}ii\hspace{.08em})$, which has inequivalent anisotropy for two triangular lattices ($\Delta_A\neq \Delta_B$), 
in a similar way to the model $\rm(\hspace{.18em}i\hspace{.18em})$ ($J_2\neq J_2^\prime$)
with some changes in details. 
We show the energy band of $h=0$, $h=2J_1S$, and $h=20J_1S$ in Figs.~\ref{fig:J1J2_berry}(a), (b) and (c), and the Berry curvature of $h=0$, $h=2J_1S$, and $h=20J_1S$ in Figs.~\ref{fig:J1J2_berry}(d), (e), and (f). When the magnetic field is zero, the Hamiltonian satisfies the effective TRS. Thus, the Berry curvature $\Omega_{n,xy}(\bm{k})$ satisfies $\Omega_{n,xy}(\bm{k})=-\Omega_{n,xy}(-\bm{k})$ as shown in Figs.~\ref{fig:J1J2_berry}(d), and the thermal Hall conductivity is zero, as in model $\rm(\hspace{.18em}i\hspace{.18em})$ ($J_2\neq J_2^\prime$). Figures ~\ref{fig:J1J2_berry}(b) and (c) show that the energy around $K^\prime$ decreases when $h\neq0$ while energy around $K$ increases, which is the same as the model $\rm(\hspace{.18em}i\hspace{.18em})$ ($J_2\neq J_2^\prime$). We show the magnetic field dependence of the Chern number and the color plot of thermal Hall conductivity in Fig.~\ref{fig:J1J2_thermal}(a). Figures~\ref{fig:J1J2_thermal}(b) shows the thermal Hall conductivity in some temperatures. Since the pattern of the inequivalence is changed, the region where the Chern number is non-zero is different between the model$\rm(\hspace{.18em}i\hspace{.18em})$ ($J_2\neq J_2^\prime$) and the model $\rm(\hspace{.08em}ii\hspace{.08em})$ ($\Delta_A\neq \Delta_B$). However, in both models, the sign of the thermal Hall conductivity corresponds to the sign of the Chern number. 

We note that our assumption for the ground state spin configuration (\ref{spin_a}) becomes not so good in the large magnetic field region.
While Figs.~\ref{fig:J1J2J3_thermal}(b) and \ref{fig:J1J2_thermal}(b) show that the thermal Hall conductivity changes dramatically about $h=21J_1S$, this region may be out of validity of our ansatz (\ref{spin_a})  because $\psi_A=\pi/2$ and $\psi_B<\pi/2$ in this region. Specifically, when $\psi_A=\pi/2$ and $\psi_B<\pi/2$, the classical energy is independent of the angle $\bm{Q}\cdot\bm{R}_i$ of the $A$ site spins and the in-plane angle of the $A$ site spins becomes arbitrary.

\section{SPIN NERNST EFFECT}

\label{nernst_sec}
In this section, we study spin current response induced by thermal gradient in frustrated honeycomb magnets. In particular, we consider a transverse response called the spin Nernst effect. 

Because the spin Hamiltonian described as Eq.~($\ref{model_ham}$) commutes with $S^z$, we can define spin current. However, a problem arises when we approximate spin Hamiltonian as a bilinear form of creation and annihilation operators of the magnon, especially when we consider noncollinear systems. 
Namely, the magnon Hamiltonian itself does not commute with spin operator $S^z$. In noncollinear systems, the spin operator $S^z$ is written by the $\bm{S}^\prime$ as Eq.~(\ref{rotation}), and $S^z$ is not a bilinear form of magnon operators unlike collinear systems. Thus, in the noncollinear systems, the commutation of $S^z$ and the magnon Hamiltonian changes the order of creation and annihilation operators of magnons.
To overcome this issue, we use a formulation of current associated with a general operator before considering the spin operator.
Specifically, we write the general operator on the magnon space as
\begin{equation*}
    O(\bm{r})=\frac{1}{2}\Psi^\dagger(\bm{r})O\Psi(\bm{r}).
\end{equation*}
Here, $\Psi$ is defined as Eq.~(\ref{def_psi}) and $O$ is the $4\times4$ matrix. Thus, we can write time differential of $O(\bm{r})$ as current $\bm{j_o}$ part and source $S_o$ part  \cite{Zyuzin2016MagnonAntiferromagnets,Cheng2016SpinAntiferromagnets,Zhang2018Spin-NernstInsulator,Park2019TopologicalAntiferromagnets,Li2020IntrinsicAntiferromagnet},
\begin{equation}
    \frac{\partial O(\bm{r})}{\partial t}=i[H,O(\bm{r})]=-\nabla\cdot \bm{j_o}+S_o,
\end{equation}
where $\bm{j_o}=\frac{1}{4}(O\sigma_3\bm{v}+\bm{v}\sigma_3 O)$, $S_o=-\frac{i}{2}(O\sigma_3 H(\bm{k})-H(\bm{k})\sigma_3O)$, and $\bm{v}=i[H(\bm{k}),\bm{r}]$.\par
Hereafter, we focus on the spin Nernst effect, and we set $O$ to be $\tilde{S}^z$, which corresponds to the magnon spin density operator given by
\begin{equation*}
    \tilde{S}^z=\begin{pmatrix}
             \sin{\psi_A} & 0 & 0 & 0 \\
             0 & \sin{\psi_B} & 0 & 0 \\
             0 & 0 & \sin{\psi_A} & 0 \\
             0 & 0 & 0 & \sin{\psi_B}\\
    \end{pmatrix}.
\end{equation*}
Using the linear response theory for $\bm{j}$, we obtain the expression for the spin Nernst effect as \cite{Li2020IntrinsicAntiferromagnet}
\begin{align}
    j_\alpha&=\alpha_{\alpha\beta}\nabla_\beta T \notag\\
    &=\frac{2k_B}{\hbar}\sum_n\int_{BZ}\frac{dk^2}{(2\pi)^2}[\Omega^{S_z}(\bm{k})]_{n,\alpha\beta}c_1(\rho(E_n(\bm{k}))\nabla_\beta T,
\end{align}
where
\begin{align}
    &[\Omega^{S_z}(\bm{k})]_{n,\alpha\beta}
    \nonumber\\
    &=
    \sum_{m\neq n}(\sigma_3)_{nn}(\sigma_3)_{mm} 
    \nonumber\\
    &\times \frac{2\textrm{Im}[\mel{\bm{t}_n(\bm{k})}{j_\alpha(\bm{k})}{\bm{t}_m(\bm{k})}\mel{\bm{t}_m(\bm{k})}{v_\beta(\bm{k})}{\bm{t}_n(\bm{k})}]}{((\sigma_3E(\bm{k}))_{nn}-(\sigma_3E(\bm{k}))_{mm})^2},
\end{align}
and
\begin{equation*}
    c_1(\rho)=(1+\rho)\log{(1+\rho)}-\rho\log{\rho}.
\end{equation*}

\begin{figure}[tb]
\includegraphics[width=0.9\linewidth]{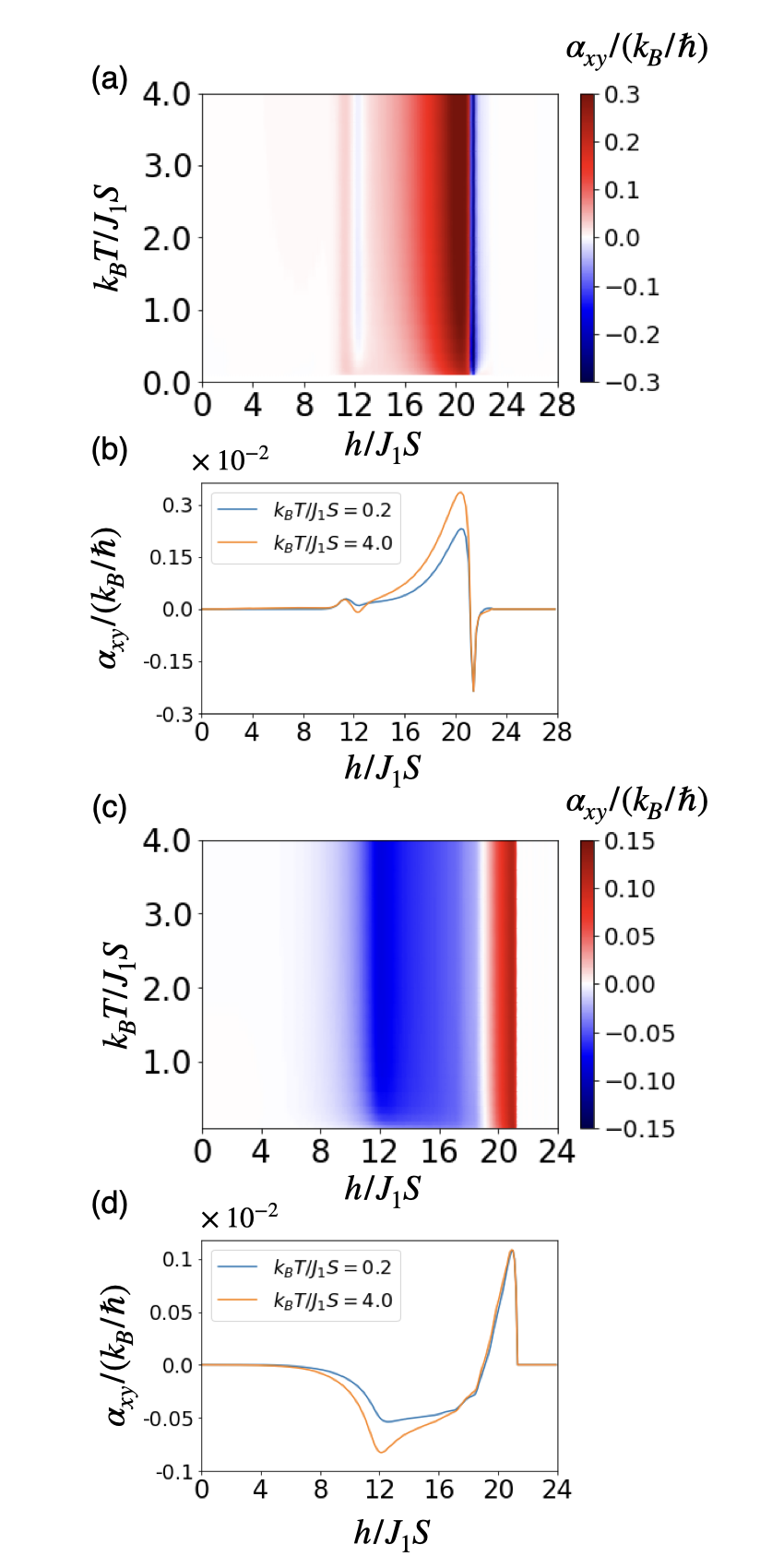}
\caption{\label{fig:J1J2_nernst} 
The spin Nernst conductivity. (a, b) The spin Nernst conductivity $\alpha_{xy}$ of the model$\rm(\hspace{.18em}i\hspace{.18em})$ with $J_1=1.0$, $J_2=2.0$, $J_2^\prime=2.2$, $\Delta_A=\Delta_B=0.05$, $S=\frac{1}{2}$. (a) $\alpha_{xy}$ as a function of the magnetic field at several temperatures, and (b) the color plot of $\alpha_{xy}$ for the model$\rm(\hspace{.18em}i\hspace{.18em})$. 
(c,d) The spin Nernst conductivity $\alpha_{xy}$ of the model$\rm(\hspace{.08em}ii\hspace{.08em})$ with $J_1=1.0$, $J_2=J_2^\prime=2.0$, $\Delta_A=0.05$, $\Delta_B=0.1$. (c) $\alpha_{xy}$ as a function of the magnetic field at several temperatures, and (d) the color plot of $\alpha_{xy}$ for the model$\rm(\hspace{.08em}ii\hspace{.08em})$.
(c) and (d) indicate that the sign of the spin Nernst conductivity  $\alpha_{xy}$ approximately corresponds to the sign of the thermal Hall conductivity.}
\end{figure}

\begin{figure}[bt]
\includegraphics[width=\linewidth]{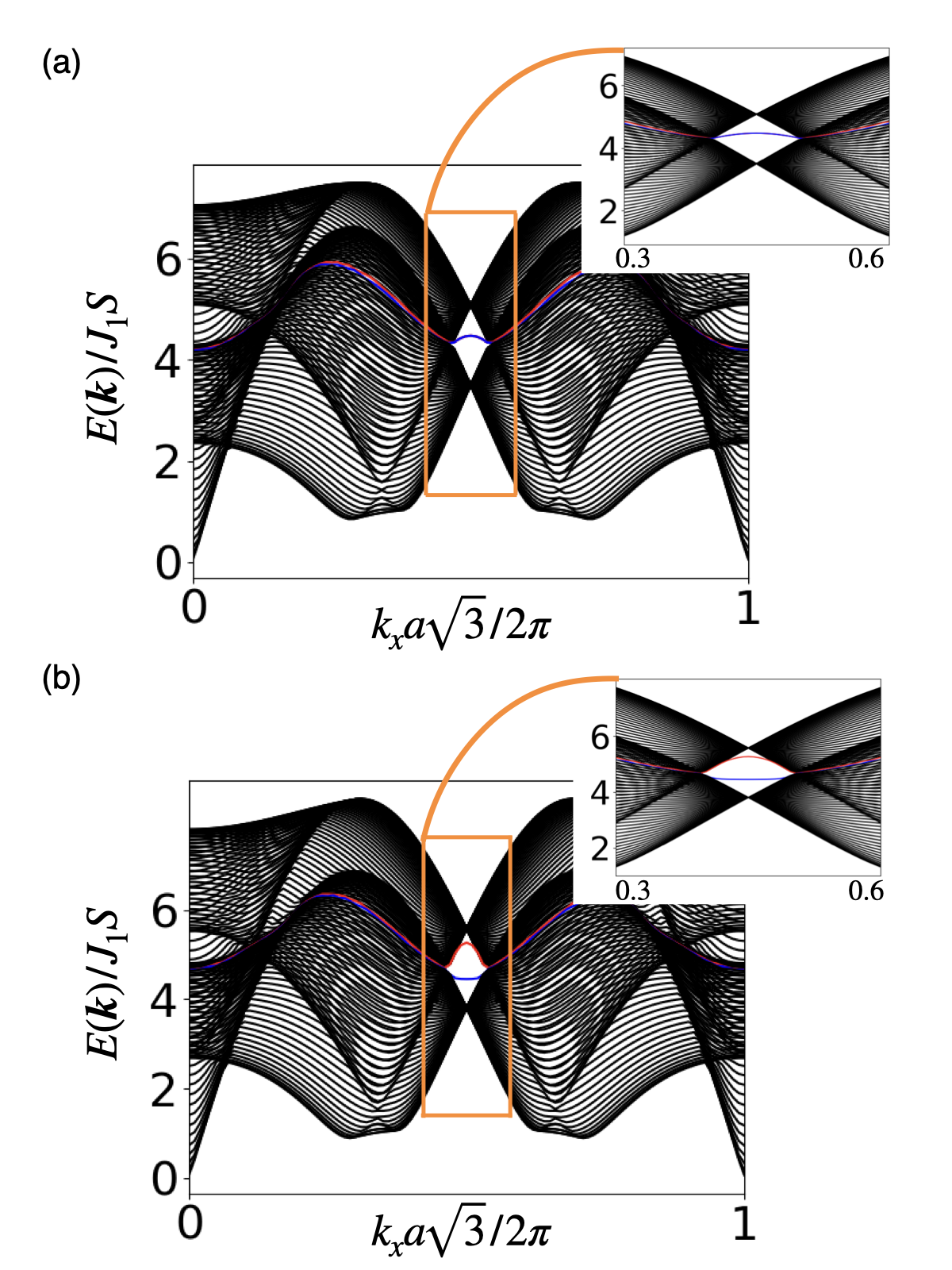}
\caption{The energy dispersion of zigzag edge with $J_1=1.0$, $h=0$. The red and blue line is edge modes. (a) The energy dispersion of the AB equivalent model, $J_2=J_2^\prime=2.0$, $\Delta_A=\Delta_B=0.05$. (b) The energy dispersion and edge modes of the model$\rm(\hspace{.08em}ii\hspace{.08em})$, $J_2=2.0$,$J_2^\prime=2.2$, $\Delta_A=\Delta_B=0.05$. Inset of (a) and (b) shows the energy dispersion around the edge. In (b) the edge modes are energetically separable, while the edge modes are degenerate in (a).}
\label{fig:Zigzag}
\end{figure}

While the formula for spin Nernst conductivity does not contain the Berry curvature, breaking of the symmetry (\ref{symmetry}) is also needed for the non-zero spin Nernst conductivity. Specifically, if $A$ sites and $B$ sites are equivalent and the symmetry (\ref{symmetry}) is satisfied, we have $Pv_\alpha(\bm{k})P=P\pdv{H(\bm{k})}{k_\alpha}P=v_\alpha^*(\bm{k})$ and $Pj_\alpha(\bm{k})P=j_\alpha^*(\bm{k})$. Thus, similarly to the Berry curvature, $\Omega^{S_z}$ must be odd in $\bm{k}$, $\Omega^{S_z}(\bm{k})_{\alpha\beta,n}=-\Omega^{S_z}(\bm{-k})_{\alpha\beta,n}$. %Therefore, the inequivalence is needed for non-zero spin Nernst conductivity. 

Figure~\ref{fig:J1J2_nernst}(a) shows the color plot of the spin Nernst conductivity for the model$\rm(\hspace{.18em}i\hspace{.18em})$ ($J_2\neq J_2^\prime$), and Fig.~\ref{fig:J1J2_nernst}(c) shows that for the model$\rm(\hspace{.08em}ii\hspace{.08em})$ ($\Delta_A\neq\Delta_B$). These figures show that the sign of the spin Nernst conductivity is approximately corresponding to the sign of the thermal Hall conductivity.
Figures~\ref{fig:J1J2_nernst}(b) and (d) show the spin Nernst conductivity for the model$\rm(\hspace{.18em}i\hspace{.18em})$ ($J_2\neq J_2^\prime$) and the model$\rm(\hspace{.08em}ii\hspace{.08em})$ ($\Delta_A\neq\Delta_B$) at several temperatures, respectively. In both cases, the spin Nernst conductivity is small when the magnetic field $h$ is small. This is because $\expval{S^z}$ is small for a small magnetic field $h$. 
Since the spin Nernst effect  in the present model requires non-zero $\expval{S^z}$,  small $h$ leads to small spin Nernst effect through its dependence on $\expval{S^z}$. 
On the other hand, when the magnetic field $h/J_1S$ is large, the behavior of the spin Nernst conductivity resembles that of the thermal Hall conductivity. While we can see a drastic change in the spin Nernst conductivity in the large magnetic field regime $h>20J_1S$, the ansatz (\ref{spin_a}) is not reasonable as we have mentioned in Sec.~III.

Finally, we show the edge modes of the magnon band. We impose a periodic boundary condition to the $x$ direction and an open boundary condition to the $y$ direction. This choice of the boundary results in the zigzag edge. 
Figure~\ref{fig:Zigzag}(a) shows the energy dispersion of the case when the $A$ sites and the $B$ sites are equivalent ($J_2=J_2^\prime$ and $\Delta_A=\Delta_B$), and Fig.~\ref{fig:Zigzag}(b) shows the case when the $A$ sites and the B sites are inequivalent ($J_2\neq J_2^\prime$). The band structures show that only when the $A$ sites and $B$ sites are inequivalent, the edge modes appearing at the opposite edges (red and blue lines) are energetically nondegenerate. Thus, when the $A$ sites and $B$ sites are inequivalent, two edge states are inequivalent and allow transverse responses of  heat and spins.

\section{Discussions}

\begin{figure}[bt]
\includegraphics[width=\linewidth]{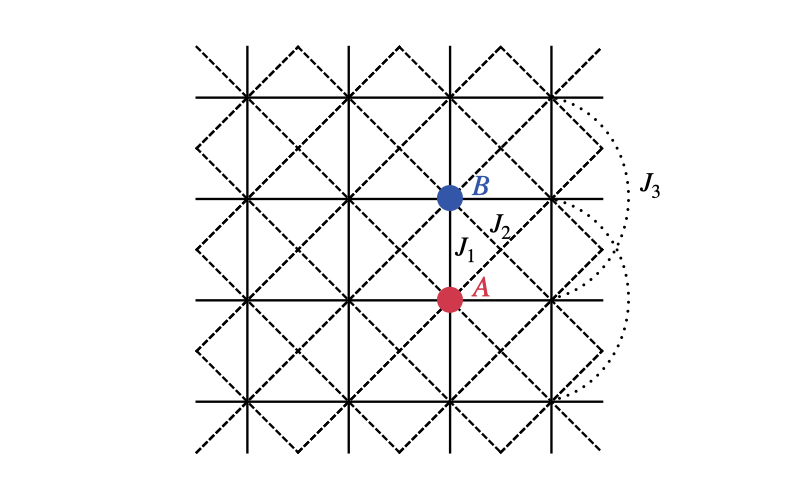}
\caption{\label{fig: square lattice}
The $J_1$-$J_2$-$J_3$ model on the square lattice. Solid, dashed and dotted lines represent $J_1$, $J_2$ and $J_3$, respectively. For visibility, the third-nearest-neighbor hopping $J_3$ is depicted only partially on the right edge.}
\end{figure}

We have established the condition of the magnon thermal Hall effect without DM interaction in terms of the symmetry of the BdG Hamiltonian. 
The symmetry argument shows that the Berry curvature is non-zero when the $A$ and $B$ sites are inequivalent. Furthermore, the canting angle from $xy$- plane $\psi_A$ and $\psi_B$ can induce the SU(2) gauge field when $\psi_A\neq\psi_B$. We also study the $J_1$-$J_2$-$J_2^\prime$ model to clarify the relation of the thermal Hall conductivity and the Chern number. 

Here, we consider materials such that $J_1$-$J_2$-$J_2^\prime$ model on the honeycomb lattice is feasible. Since we set the parameter $J_2>J_1$ in Sec.~\ref{J1J2J3_sec}, we can regard $J_1$-$J_2$-$J_2^\prime$ model on the honeycomb lattice as the bilayer triangular lattice. One of the candidate materials of antiferromagnetic Heisenberg model on the triangular lattice is \ce{Ba3XSb2O9} (\ce{X} = \ce{Mn}, \ce{Co}, and \ce{Ni}) \cite{Doi2004StructuralNi,Shirata2012ExperimentalAntiferromagnet,Zhou2012SuccessiveBa3CoSb2O9,Susuki2013MagnetizationBa3CoSb2O9,Quirion2015MagneticMeasurements,Ma2016StaticBa3CoSb2O9,Maksimov2016Field-inducedAntiferromagnets}. The materials \ce{Ba3XSb2O9} contains stacked triangular lattice, but these layers are equivalent. Thus, we need to add inequivalence to each layer, for example, by adding an electric field in the direction of $c$-axis. Another candidate material is TMD. In particular, numerical calculations suggest that the ground state of \ce{VX2} and \ce{MnX2} (\ce{X} = \ce{Cr}, \ce{Br}, and \ce{I}) has a $120^\circ$ antiferromagnetic spin configurations \cite{Li2020HighMonolayers}. Thus, we may create a $J_1$-$J_2$-$J_2^\prime$ model by heterostacking \ce{VX2} and \ce{MnX2}.

By using the parameter of \ce{Ba3CoSb2O9}, we estimate the thermal Hall conductivity in units of W/Km. Inter-layer distance $d\sim 15~\AA$ and the intra-layer coupling $J_2/k_B \sim 18 K~$\cite{Doi2004StructuralNi,Shirata2012ExperimentalAntiferromagnet}. By using these parameters and assume $J_1\sim J_2/2$ and $k_BT\sim J_1\sim 9~k_BT$, the unit of the thermal Hall conductivity $\kappa_{xy}/(k_B/\hbar)$ approximately corresponds to $\kappa_{xy}\sim$0.01~W/Km. Therefore, the order of the thermal Hall conductivity is $10^{-2}$~W/Km in our models. 
This value is comparable to that for kagome antiferromagnets with in-plane DM interactions that was studied in Ref.~\onlinecite{Laurell2018MagnonInteractions}.

Finally, we comment on models other than the $J_1$-$J_2$-$J_2^\prime$ model on the honeycomb lattice, where the thermal Hall effect may occur without DM interaction. One candidate is $J_1$-$J_2$-$J_3$ model on the square lattice 
(as illustrated in Fig.~\ref{fig: square lattice}) 
whose classical ground state exhibits a spiral phase \cite{Moreo1990IncommensurateModels,Chubukov1991First-orderAntiferromagnets,Rastelli1992NonlinearPhase,Ferrer1993Spin-liquidLattice,Ceccatto1993NonclassicalAntiferromagnets,Reuther2011QuantumModel}. The square lattice is a bipartite lattice and we can define $A$ sites and $B$ sites. To support nonzero thermal Hall response, the inequivalence of two sublattices can be introduced by changing the magnetic anisotropy or next-nearest-neighbor hopping of $A$ sites and $B$ sites. 
This leads to the Hamiltonian written as
\begin{align*}
    H=&\sum_{\langle i,j\rangle}J_1 \bm{S}_i\cdot\bm{S}_j+\sum_{\langle\langle i,j\rangle\rangle\in A}J_2 \bm{S}_i\cdot\bm{S}_j+\sum_{\langle\langle i,j\rangle\rangle\in B}J^\prime_2 \bm{S}_i\cdot\bm{S}_j\\
    &+\sum_{\langle\langle\langle i,j\rangle\rangle\rangle\in A}J_3 \bm{S}_i\cdot\bm{S}_j+\sum_{\langle\langle\langle i,j\rangle\rangle\rangle\in B}J^\prime_3 \bm{S}_i\cdot\bm{S}_j\\
    &+\sum_{i\in A}\Delta_A (S^z_i)^2+\sum_{i\in B}\Delta_B (S^z_i)^2+h\sum_i S^z_i,
\end{align*}
where $\sum_{\langle\langle\langle i,j\rangle\rangle\rangle}$ means that sum over third-nearest-neighbor of the square lattice. In this model, $J_2\neq J_2^\prime$ or $J_3\neq J_3^\prime$ or $\Delta_A\neq \Delta_B$ will support the non-zero Berry curvature and thermal Hall responses.

\begin{acknowledgments}
We thank Shuichi Murakami for fruitful discussions.
This work was supported by 
JSPS KAKENHI Grant 20K14407 (S.K.), 
JST CREST (Grant No. JPMJCR19T3) (S.K., T.M.),
and JST PRESTO (Grant No. JPMJPR19L9) (T.M.).
K.F. was supported by the Forefront Physics and Mathematics program to drive transformation (FoPM)
\end{acknowledgments}

\appendix

\section{Details of the magnon Hamiltonian in the spiral phase}
In this section, we show details of the calculation of the magnon Hamiltonian  ($\ref{generalisedmodel}$) in the spiral phase, and identify the symmetry-breaking interactions based on the condition Eq.~(\ref{eq:general-condition}) leading to the effective $PT$ symmetry.

First, we rewrite the spin Hamiltonian ($\ref{generalisedmodel}$) in the rotated spin coordinate ($\ref{rotation}$) as
\begin{align}
    H=&\sum_{i,j}J_{\alpha\beta}\left[(\sin{\psi_i}\sin{\psi_j}\cos{\theta_{ij}}+\cos{\psi_i}\cos{\psi_j})S^{\prime x}_iS^{\prime x}_j \right.\notag\\
    &+\cos{\theta_{ij}}S^{\prime y}_iS^{\prime y}_j \notag \\
    &+(\cos{\psi_i}\cos{\psi_j}\cos{\theta_{ij}}+\sin{\psi_i}\sin{\psi_j})S^{\prime z}_iS^{\prime z}_j \notag \\
    &\left.-\sin{\psi_i}\sin{\theta_{ij}}S^{\prime x}_iS^{\prime y}_j+\sin{\psi_j}\sin{\theta_{ij}}S^{\prime y}_iS^{\prime x}_j\right] \notag \\
    &+\sum_i\left[h\sin{\psi_\alpha}S^{\prime z}_i+\Delta_\alpha\left(\cos^2{\psi_i}(S^{\prime x}_i)^2+\sin^2{\psi_i}(S^{\prime z}_i)^2\right)\right] \notag \\
    &+(S^{\prime x}_iS^{\prime z}_j \textrm{~and~} S^{\prime y}_iS^{\prime z}_j \textrm{~terms}). \label{canted_ham}
\end{align}
where $\alpha,\beta=A,B$ denote the sublattices to which $i$ and $j$ sites belong, respectively, and $\theta_{ij}=\bm{Q}\cdot\bm{R}_j+\phi_j-\bm{Q}\cdot\bm{R}_i-\phi_i$.
To this Hamiltonian ($\ref{canted_ham}$), we apply the HP transformation ($\ref{HPtrans}$), and obtain the magnon Hamiltonian in the form of
\begin{equation}
H(\bm{k})=H_0+\sum_{\bm{R}}H(\bm{k},\bm{R},\bm{r}).
\end{equation}
Here, $H_0$ consists of the local terms, i.e., the easy-axis anisotropy and the Zeeman term, while 
$H(\bm{k},\bm{R},\bm{r})$ represents the Heisenberg interaction part. 
The vector $\bm{R}$ denotes the distance between centers of unit cells $\bm{R}_j-\bm{R}_i$, and the summation is taken over all the unit cells (with fixing $\bm{R}_i$ at the origin).
The vector $\bm{r}$ is a short-hand notation for the distance between $i$ site and $j$ site, and takes $\bm{r}=\bm{R}$ for the diagonal part (e.g. $\Xi^0$ and $\Xi^z$) and $\bm{r}=\bm{R}+\bm{\delta}$ for the offdiagonal part (e.g. $\Xi^x$ and $\Xi^y$) of $2\times2$ blocks in the following, where $\bm{\delta}$ is defined as a distance from the $A$ site to the $B$ site in the same unit cell.

Let us write the magnon Hamiltonian $H(\bm{k},\bm{R},\bm{r})$ as
\begin{equation*}
H(\bm{k},\bm{R},\bm{r})=
\begin{pmatrix}
          \Xi(\bm{k},\bm{R},\bm{r}) & \Pi(\bm{k},\bm{R},\bm{r}) \\
          \Pi^*(-\bm{k},\bm{R},\bm{r}) & \Xi^*(-\bm{k},\bm{R},\bm{r}) 
        \end{pmatrix}.
\end{equation*}
Using Pauli matrices, we expand $\Xi(\bm{k},\bm{R},\bm{r})$ and $\Pi(\bm{k},\bm{R},\bm{r})$ as
\begin{align*}
    \Xi(\bm{k},\bm{R},\bm{r})=&\Xi^0(\bm{k},\bm{R},\bm{r})\sigma_0+\Xi^x(\bm{k},\bm{R},\bm{r})\sigma_x\\
   &+\Xi^y(\bm{k},\bm{R},\bm{r})\sigma_y+\Xi^z(\bm{k},\bm{R},\bm{r})\sigma_z, \\
    \Pi(\bm{k},\bm{R},\bm{r})=&\Pi^0(\bm{k},\bm{R},\bm{r})\sigma_0+\Pi^x(\bm{k},\bm{R},\bm{r})\sigma_x\\
   &+\Pi^y(\bm{k},\bm{R},\bm{r})\sigma_y+\Pi^z(\bm{k},\bm{R},\bm{r})\sigma_z.
\end{align*}
From the symmetry analysis, we show that non-zero $\Xi^z(\bm{k},\bm{R},\bm{r})$, $\Pi^z(\bm{k},\bm{R},\bm{r})$, or $\Im{\Pi^i(\bm{k},\bm{R},\bm{r})}$ may leads to the non-zero Berry curvature (see Eq.~(\ref{eq:general-condition})). Here, each coefficient of the Pauli matrices for $\Xi$ is given as follows:
\begin{subequations}
\label{bdg_all}
\begin{align}
    &\Xi^0(\bm{k},\bm{R},\bm{r})=\nonumber\notag\\
    &-SJ_{AB}(\bm{r})[\cos{(\bm{R}\cdot\bm{Q}+\phi)}\notag\\
    &+\cos{\psi_A}\cos{\psi_B}+\sin{\psi_A}\sin{\psi_B}]\notag\\
    &-\frac{S}{2}J_{AA}(\bm{r})[\sin{(\bm{R}\cdot\bm{Q})}\sin{\psi_A}\sin{(\bm{k}\cdot\bm{r})}\notag\\
    &+(\cos^2{\psi_A}\cos{(\bm{R}\cdot\bm{Q})}+\sin^2{\psi_A})\notag\\
    &-\frac{1}{2}(\cos{(\bm{R}\cdot\bm{Q})}(1+\sin^2{\psi_A})+\cos^2{\psi_A})\cos{(\bm{k}\cdot\bm{r})}] \notag\\
    &-\frac{S}{2}J_{BB}(\bm{r})[\sin{(\bm{R}\cdot\bm{Q})}\sin{\psi_B}\sin{(\bm{k}\cdot\bm{r})}\notag\\
    &+(\cos^2{\psi_B}\cos{(\bm{R}\cdot\bm{Q})}+\sin^2{\psi_B})\notag\\
    &-\frac{1}{2}(\cos{(\bm{R}\cdot\bm{Q})}(1+\sin^2{\psi_B})+\cos^2{\psi_B})\cos{(\bm{k}\cdot\bm{r})}], \label{xi_0}
\end{align}
\begin{align}
    \Xi^x(\bm{k},\bm{R},\bm{r})=&\frac{S}{2}J_{AB}(\bm{r})[\{\cos{(\bm{R}\cdot\bm{Q}+\phi)}(1+\sin{\psi_A}\sin{\psi_B})\notag\\
    &+\cos{\psi_A}\cos{\psi_B}\}\cos({\bm{k}\cdot\bm{r}})\notag\\
    &-\sin{(\bm{R}\cdot\bm{Q}+\phi)}(\sin{\psi_A}+\sin{\psi_B})\sin({\bm{k}\cdot\bm{r}})], \label{xi_x}
\end{align}
\begin{align}
    \Xi^y(\bm{k},\bm{R},\bm{r})=&-\frac{S}{2}J_{AB}(\bm{r})[\{\cos{(\bm{R}\cdot\bm{Q}+\phi)}(1+\sin{\psi_A}\sin{\psi_B})\notag\\
    &+\cos{\psi_A}\cos{\psi_B}\}\sin{\bm{k}\cdot\bm{r}}\notag\\
    &+\sin{(\bm{R}\cdot\bm{Q}+\phi)}(\sin{\psi_A}+\sin{\psi_B})\cos({\bm{k}\cdot\bm{r}})], \label{xi_y}
\end{align}
\begin{align}
    \Xi^z(\bm{k},\bm{R},\bm{r})=&-\frac{S}{2}J_{AA}(\bm{r})[\sin{(\bm{R}\cdot\bm{Q})}\sin{\psi_A}\sin{(\bm{k}\cdot\bm{r})}\notag\\
    &+(\cos^2{\psi_A}\cos{(\bm{R}\cdot\bm{Q})}+\sin^2{\psi_A})\notag\\
    &-\frac{1}{2}(\cos{(\bm{R}\cdot\bm{Q})}(1+\sin^2{\psi_A})+\cos^2{\psi_A})\cos{(\bm{k}\cdot\bm{r})}] \notag\\
    &+\frac{S}{2}J_{BB}(\bm{r})[\sin{(\bm{R}\cdot\bm{Q})}\sin{\psi_B}\sin{(\bm{k}\cdot\bm{r})}\notag\\
    &+(\cos^2{\psi_B}\cos{(\bm{R}\cdot\bm{Q})}+\sin^2{\psi_B})\notag\\
    &-\frac{1}{2}(\cos{(\bm{R}\cdot\bm{Q})}(1+\sin^2{\psi_B})+\cos^2{\psi_B})\cos{(\bm{k}\cdot\bm{r})}], \label{xi_z}
\end{align}
where $\Xi^z(\bm{k},\bm{R},\bm{r})=0$ if $J_{AA}=J_{BB}$ and $\psi_A=\psi_B$. 
Similarly, the coefficients for $\Pi(\bm{k},\bm{R},\bm{r})$ are written as
\begin{align}
    \Pi^0(\bm{k},\bm{R},\bm{r})=&\frac{S}{4}(\cos{(\bm{R}\cdot\bm{Q})}-1)\cos{(\bm{k}\cdot\bm{r})}\notag\\
    &(J_{AA}(\bm{r})\cos^2{\psi_A}+J_{BB}(\bm{r})\cos^2{\psi_B}),\label{pi_0}
\end{align}
\begin{align}
    \Pi^x(\bm{k},\bm{R},\bm{r})=&\frac{S}{2}J_{AB}(\bm{r})[(\cos({\bm{R}\cdot\bm{Q}}+\phi)(\sin{\psi_A}\sin{\psi_B}-1)\notag\\
    &+\cos{\psi_A}\cos{\psi_B})\cos{\bm{k}\cdot\bm{r}}\notag\\&
    -i\sin({\bm{R}\cdot\bm{Q}}+\phi)(\sin{\psi_A}-\sin{\psi_B})\cos{(\bm{k}\cdot\bm{r})}], \label{pi_x}
\end{align}
\begin{align}
    \Pi^y(\bm{k},\bm{R},\bm{r})=&-\frac{S}{2}J_{AB}(\bm{r})[(\cos({\bm{R}\cdot\bm{Q}}+\phi)(\sin{\psi_A}\sin{\psi_B}-1)\notag\\
    &+\cos{\psi_A}\cos{\psi_B})\sin{\bm{k}\cdot\bm{r}}\notag\\&
    -i\sin({\bm{R}\cdot\bm{Q}}+\phi)(\sin{\psi_A}-\sin{\psi_B})\sin{(\bm{k}\cdot\bm{r})}], \label{pi_y}
\end{align}
\begin{align}
    \Pi^z(\bm{k},\bm{R},\bm{r})=&\frac{S}{4}(\cos{(\bm{R}\cdot\bm{Q})}-1)\cos{(\bm{k}\cdot\bm{r})}\notag\\
    &(J_{AA}(\bm{r})\cos^2{\psi_A}-J_{BB}(\bm{r})\cos^2{\psi_B}).\label{pi_z}
\end{align}
\end{subequations}
Here we find that 
$\Im{\Pi^x(\bm{k},\bm{R},\bm{r})}$ and $\Im{\Pi^y(\bm{k},\bm{R},\bm{r})}$ depend on $\sin{\psi_A}-\sin{\psi_B}$ and $\sin{(\bm{R}\cdot\bm{Q}+\phi)}$. Thus, these are non-zero only if $\psi_A\neq\psi_B$ and
$\sin{(\bm{R}\cdot\bm{Q}+\phi)}\neq0$. 
The expression for $\Pi^z$ indicates that $\Pi^z(\bm{k},\bm{R},\bm{r})$ vanishes if $\psi_A=\psi_B$ and $J_{AA}=J_{BB}$ is satisfied.

Next, we write $H_0(\bm{k})$ as
\begin{equation*}
H_0=
\begin{pmatrix}
          \Xi_0 & \Pi_0 \\
          \Pi_0^* & \Xi_0^* 
        \end{pmatrix},
\end{equation*}
and expand $\Xi_0$ and $\Pi_0$ as
\begin{align*}
    \Xi_0=&\Xi_0^0\sigma_0+\Xi_0^x\sigma_x+\Xi_0^y\sigma_y+\Xi_0^z\sigma_z, \notag\\
    \Pi_0=&\Pi_0^0\sigma_0+\Pi_0^x\sigma_x+\Pi_0^y\sigma_y+\Pi_0^z\sigma_z, 
\end{align*}
where each coefficient of the Pauli matrices is given as follows:
\begin{subequations}
\begin{align}
    \Xi_0^0=&\frac{S}{2}(\Delta_A(1-3\sin^2{\psi_A})+\Delta_B(1-3\sin^2{\psi_B}))\notag\\
    &+\frac{h}{2}(\sin{\psi_A}+\sin{\psi_B}),
\end{align}
\begin{align}
    \Xi_0^z=&\frac{S}{2}(\Delta_A(1-3\sin^2{\psi_A})-\Delta_B(1-3\sin^2{\psi_B}))\notag\\
    &+\frac{h}{2}(\sin{\psi_A}-\sin{\psi_B}),
\end{align}
\begin{align}
    \Pi_0^0=\frac{S}{2}(\Delta_A\cos^2{\psi_A}+\Delta_B\cos^2{\psi_B}),
\end{align}
\begin{align}
    \Pi_0^z=\frac{S}{2}(\Delta_A\cos^2{\psi_A}-\Delta_B\cos^2{\psi_B}),\label{pi_z0}
\end{align}
\begin{align}
    \Xi_0^x=\Xi_0^y=\Pi_0^x=\Pi_0^y=0.
\end{align}
\end{subequations}
Here $\Xi_0^z$ is zero if $\psi_A=\psi_B$. The above terms imply that if $A$ sites and $B$ sites are equivalent (i.e., $J_{AA}=J_{BB}$, $\Delta_A=\Delta_B$, and $\theta_A=\theta_B$), $\Xi^z(\bm{k},\bm{R},\bm{r})=\Pi^z(\bm{k},\bm{R},\bm{r})=\Im{\Pi^i(\bm{k},\bm{R},\bm{r})}= \Xi^z_0=\Pi^z_0=0$. Thus, the Hamiltonian satisfies the symmetry~(\ref{symmetry}) leading to the vanishing Berry curvature. 

Now, we consider the effective TRS $H(\bm{k})=H^*(-\bm{k})$. In the above expressions, we can see that terms proportional to  $\sin\bm{k}\cdot\bm{r}$ for the real part and $\cos\bm{k}\cdot\bm{r}$ for the imaginary part lead to the broken effective TRS. These terms are proportional to $\sin{(\bm{R}\cdot\bm{Q})}\sin{\psi_i}$ [see Eqs.~(\ref{xi_0},\ref{xi_z})], $\sin{(\bm{R}\cdot\bm{Q}+\phi)}(\sin{\psi_A}+\sin{\psi_B})$ [see Eqs.~(\ref{xi_x},\ref{xi_y})], and $\sin{(\bm{R}\cdot\bm{Q}+\phi)}(\sin{\psi_A}-\sin{\psi_B})$ [see Eqs.~(\ref{pi_x},\ref{pi_y})]. Thus, $\bm{R}\cdot\bm{Q}=0$ and $\bm{R}\cdot\bm{Q}+\phi=0$ for all $\bm{R}$ or $\sin{\psi_A}=\sin{\psi_B}=0$ support the effective TRS.
This condition is independent from that for the effective $PT$ symmetry, and leads to the vanishing thermal Hall conductivity even if we have the nonzero Berry curvature. For instance, 
when $\bm{R}\cdot\bm{Q}+\phi=0$ for all $\bm{R}$ or $\sin{\psi_A}=\sin{\psi_B}=0$, we can still break the effective $PT$ symmetry with $J_{AA}\neq J_{BB}$ (or $\Delta_A\neq\Delta_B$), via non-zero $\Xi^z(\bm{k},\bm{R},\bm{r})$ (non-zero $\Xi^z_0$). 

Above expressions are applicable to general BdG Hamiltonians of $AB$ sublattice systems in the spiral phase. Once we consider the specific model, we assign a concrete value to $J_{\alpha\beta}$; For example, in $J_1$-$J_2$-$J_2^\prime$ model, $J_{AB}$ with $|\bm{r}|=a$ is $J_1$ for the nearest-neighbor $i,j$ sites, and $J_{AA}$ ($J_{BB}$) with $|\bm{r}|=\sqrt{3}a$ is $J_2$ ($J_2^\prime$) for the next-nearest-neighbor sites. 

Furthermore, we calculate the part of the magnon Hamiltonian for the DM interaction of the following form
\begin{equation*}
H_{\text{DM}}=\sum_{i,j}D_{\alpha\beta}(\bm{S}_i\times \bm{S}_j)_z.
\end{equation*}
Here, we again consider the spiral phase and rewrite $H_{\text{DM}}$ by $\bm{S}^\prime$, which results in
\begin{align*}
H_{\text{DM}}=\sum_{i,j}D_{\alpha\beta}[&\cos{\theta_{ij}}(\sin{\psi_i}S^{\prime x}_iS^{\prime y}_j-\sin{\psi_j}S^{\prime x}_jS^{\prime y}_i)\\
&+\sin{\theta_{ij}}\{\sin{\psi_i}\sin{\psi_j}S^{\prime x}_iS^{\prime x}_j\\
&+S^{\prime y}_iS^{\prime y}_j+\cos{\psi_i}\cos{\psi_j}S^{\prime z}_iS^{\prime z}_j\}].
\end{align*}
Using the HP transformation and the Fourier transformation, we obtain the DM interaction of magnons. The DM interaction term is also $4\times4$ BdG matrix of the form
\begin{equation*}
H_{\text{DM}}(\bm{k},\bm{R},\bm{r})=
\begin{pmatrix}
          \Xi_{\text{DM}}(\bm{k},\bm{R},\bm{r}) & \Pi_{\text{DM}}(\bm{k},\bm{R},\bm{r}) \\
          \Pi_{\text{DM}}^*(-\bm{k},\bm{R},\bm{r}) & \Xi_{\text{DM}}^*(-\bm{k},\bm{R},\bm{r}) \\
        \end{pmatrix}.
\end{equation*}
Then, using Pauli matrices, we expand $\Xi_{\text{DM}}(\bm{k},\bm{R},\bm{r})$ and $\Pi_{\text{DM}}(\bm{k},\bm{R},\bm{r})$ as
\begin{align*}
    \Xi_{\text{DM}}(\bm{k},\bm{R},\bm{r})=&\Xi_{\text{DM}}^0(\bm{k},\bm{R},\bm{r})\sigma_0+\Xi_{\text{DM}}^x(\bm{k},\bm{R},\bm{r})\sigma_x\\
   &+\Xi_{\text{DM}}^y(\bm{k},\bm{R},\bm{r})\sigma_y+\Xi_{\text{DM}}^z(\bm{k},\bm{R},\bm{r})\sigma_z, \\
    \Pi_{\text{DM}}(\bm{k},\bm{R},\bm{r})=&\Pi_{\text{DM}}^0(\bm{k},\bm{R},\bm{r})\sigma_0+\Pi_{\text{DM}}^x(\bm{k},\bm{R},\bm{r})\sigma_x\\
   &+\Pi_{\text{DM}}^y(\bm{k},\bm{R},\bm{r})\sigma_y+\Pi_{\text{DM}}^z(\bm{k},\bm{R},\bm{r})\sigma_z,
\end{align*}
where each coefficient of the Pauli matrices is as follows:
\begin{subequations}
\begin{align}
    &\Xi_{\text{DM}}^0(\bm{k},\bm{R},\bm{r})=\nonumber \notag\\
    &-SD_{AB}(\bm{r})\sin{(\bm{R}\cdot\bm{Q}+\phi)}\cos{\psi_A}\cos{\psi_B} \notag\\
    &+\frac{S}{2}D_{AA}(\bm{r})[\cos{(\bm{R}\cdot\bm{Q})}\sin{\psi_A}\sin{(\bm{k}\cdot\bm{r})}-\cos^2{\psi_A}\sin{(\bm{R}\cdot\bm{Q})} \notag\\
    &+\frac{1}{2}\sin{(\bm{R}\cdot\bm{Q})}(1+\sin^2{\psi_A})\cos{(\bm{k}\cdot\bm{r})}] \notag\\
    &+\frac{S}{2}D_{BB}(\bm{r})[\cos{(\bm{R}\cdot\bm{Q})}\sin{\psi_B}\sin{(\bm{k}\cdot\bm{r})}-\cos^2{\psi_B}\sin{(\bm{R}\cdot\bm{Q})} \notag\\
    &+\frac{1}{2}\sin{(\bm{R}\cdot\bm{Q})}(1+\sin^2{\psi_B})\cos{(\bm{k}\cdot\bm{r})}],
\end{align}
\begin{align}
    &\Xi_{\text{DM}}^x(\bm{k},\bm{R},\bm{r})=\nonumber\\
    &\frac{S}{2}D_{AB}(\bm{r})[\sin{(\bm{R}\cdot\bm{Q}+\phi)}(1+\sin{\psi_A}\sin{\psi_B})\cos({\bm{k}\cdot\bm{r}}) \notag\\
    &+\cos{(\bm{R}\cdot\bm{Q}+\phi)}(\sin{\psi_A}+\sin{\psi_B})\sin({\bm{k}\cdot\bm{r}})],
\end{align}
\begin{align}
    &\Xi_{\text{DM}}^y(\bm{k},\bm{R},\bm{r})=\nonumber\\
    &\frac{S}{2}D_{AB}(\bm{r})[\sin{(\bm{R}\cdot\bm{Q}+\phi)}(1+\sin{\psi_A}\sin{\psi_B})\sin({\bm{k}\cdot\bm{r}}) \notag\\
    &-\cos{(\bm{R}\cdot\bm{Q}+\phi)}(\sin{\psi_A}+\sin{\psi_B})\cos({\bm{k}\cdot\bm{r}})],
\end{align}
\begin{align}
    \Xi_{\text{DM}}^z(\bm{k},\bm{R},\bm{r})=&\frac{S}{2}D_{AA}(\bm{r})[\cos{(\bm{R}\cdot\bm{Q})}\sin{\psi_A}\sin{(\bm{k}\cdot\bm{r})} \notag\\
    &-\cos^2{\psi_A}\sin{(\bm{R}\cdot\bm{Q})} \notag\\
    &+\frac{1}{2}\sin{(\bm{R}\cdot\bm{Q})}(1+\sin^2{\psi_A})\cos{(\bm{k}\cdot\bm{r})}] \notag\\
    &-\frac{S}{2}D_{BB}(\bm{r})[\cos{(\bm{R}\cdot\bm{Q})}\sin{\psi_B}\sin{(\bm{k}\cdot\bm{r})} \notag\\
    &-\cos^2{\psi_B}\sin{(\bm{R}\cdot\bm{Q})} \notag\\
    &+\frac{1}{2}\sin{(\bm{R}\cdot\bm{Q})}(1+\sin^2{\psi_B})\cos{(\bm{k}\cdot\bm{r})}], 
\end{align}
\begin{align}
    \Pi_{\text{DM}}^0(\bm{k},\bm{R},\bm{r})=&-\frac{S}{4}\sin{(\bm{R}\cdot\bm{Q})}\cos{(\bm{k}\cdot\bm{r})} \notag\\
    &(D_{AA}(\bm{r})\cos^2{\psi_A}+D_{BB}(\bm{r})\cos^2{\psi_B}),
\end{align}
\begin{align}
    \Pi_{\text{DM}}^x(\bm{k},\bm{R},\bm{r})=&\frac{S}{2}D_{AB}(\bm{r})[\sin({\bm{R}\cdot\bm{Q}}+\phi)(\sin{\psi_A}\sin{\psi_B}-1) \notag\\
    &\cos{\bm{k}\cdot\bm{r}} \notag\\&
    +i\cos({\bm{R}\cdot\bm{Q}}+\phi)(\sin{\psi_B}-\sin{\psi_A})\cos{(\bm{k}\cdot\bm{r})}],
\end{align}
\begin{align}
    \Pi_{\text{DM}}^y(\bm{k},\bm{R},\bm{r})=&\frac{S}{2}D_{AB}(\bm{r})[\sin({\bm{R}\cdot\bm{Q}}+\phi)(\sin{\psi_A}\sin{\psi_B}-1) \notag\\
    &\sin{\bm{k}\cdot\bm{r}} \notag\\&
    +i\cos({\bm{R}\cdot\bm{Q}}+\phi)(\sin{\psi_B}-\sin{\psi_A})\sin{(\bm{k}\cdot\bm{r})}],
\end{align}
\begin{align}
    \Pi_{\text{DM}}^z(\bm{k},\bm{R},\bm{r})=&-\frac{S}{4}\sin{(\bm{R}\cdot\bm{Q})}\cos{(\bm{k}\cdot\bm{r})}\notag\\
    &(D_{AA}(\bm{r})\cos^2{\psi_A}-D_{BB}(\bm{r})\cos^2{\psi_B}).
\end{align}
\end{subequations}
Here $\Xi_{\text{DM}}^z(\bm{k},\bm{R},\bm{r})$ and $\Pi_{\text{DM}}^z$ are non-zero if $\psi_A\neq\psi_B$ or $D_{AA}\neq D_{BB}$, while
$\Im{\Pi_{\text{DM}}^x(\bm{k},\bm{R},\bm{r})}$ and $\Im{\Pi_{\text{DM}}^y(\bm{k},\bm{R},\bm{r})}$ are non-zero only if $\psi_A\neq\psi_B$. We note that $\Im{\Pi_{\text{DM}}^x(\bm{k},\bm{R},\bm{r})}$ and $\Im{\Pi_{\text{DM}}^y(\bm{k},\bm{R},\bm{r})}$ are non-zero even if $\sin({\bm{R}\cdot\bm{Q}}+\phi)$ is zero, in contrast to the Heisenberg term [$\Im{\Pi^x(\bm{k},\bm{R},\bm{r})}$ and $\Im{\Pi^y(\bm{k},\bm{R},\bm{r})}$].

From these terms, we can see that $D_{AA}\neq D_{BB}$ break the symmetry (\ref{symmetry}). Furthermore, $\Pi_{\text{DM}}^z(\bm{k},\bm{R},\bm{r})$ contains the $\sin{(\bm{k}\cdot\bm{r})}$ terms, which can be non-zero even in the collinear phase where $\psi_A=\pm\psi_B=\pi/2$ and $\bm{Q}=0$. In these cases, the DM interaction acts as the virtual magnetic field and generate non-zero Berry curvature (see Sec.~\ref{SU(2)sec}).

%\nocite{*}

\bibliographystyle{apsrev4-1}
\bibliography{references}% Produces the bibliography via BibTeX.

\begin{thebibliography}{55}%
\makeatletter
\providecommand \@ifxundefined [1]{%
 \@ifx{#1\undefined}
}%
\providecommand \@ifnum [1]{%
 \ifnum #1\expandafter \@firstoftwo
 \else \expandafter \@secondoftwo
 \fi
}%
\providecommand \@ifx [1]{%
 \ifx #1\expandafter \@firstoftwo
 \else \expandafter \@secondoftwo
 \fi
}%
\providecommand \natexlab [1]{#1}%
\providecommand \enquote  [1]{``#1''}%
\providecommand \bibnamefont  [1]{#1}%
\providecommand \bibfnamefont [1]{#1}%
\providecommand \citenamefont [1]{#1}%
\providecommand \href@noop [0]{\@secondoftwo}%
\providecommand \href [0]{\begingroup \@sanitize@url \@href}%
\providecommand \@href[1]{\@@startlink{#1}\@@href}%
\providecommand \@@href[1]{\endgroup#1\@@endlink}%
\providecommand \@sanitize@url [0]{\catcode `\\12\catcode `\$12\catcode
  `\&12\catcode `\#12\catcode `\^12\catcode `\_12\catcode `\%12\relax}%
\providecommand \@@startlink[1]{}%
\providecommand \@@endlink[0]{}%
\providecommand \url  [0]{\begingroup\@sanitize@url \@url }%
\providecommand \@url [1]{\endgroup\@href {#1}{\urlprefix }}%
\providecommand \urlprefix  [0]{URL }%
\providecommand \Eprint [0]{\href }%
\providecommand \doibase [0]{http://dx.doi.org/}%
\providecommand \selectlanguage [0]{\@gobble}%
\providecommand \bibinfo  [0]{\@secondoftwo}%
\providecommand \bibfield  [0]{\@secondoftwo}%
\providecommand \translation [1]{[#1]}%
\providecommand \BibitemOpen [0]{}%
\providecommand \bibitemStop [0]{}%
\providecommand \bibitemNoStop [0]{.\EOS\space}%
\providecommand \EOS [0]{\spacefactor3000\relax}%
\providecommand \BibitemShut  [1]{\csname bibitem#1\endcsname}%
\let\auto@bib@innerbib\@empty
%</preamble>
\bibitem [{\citenamefont {Chumak}\ \emph {et~al.}(2015)\citenamefont {Chumak},
  \citenamefont {Vasyuchka}, \citenamefont {Serga},\ and\ \citenamefont
  {Hillebrands}}]{Chumak2015MagnonSpintronics}%
  \BibitemOpen
  \bibfield  {author} {\bibinfo {author} {\bibfnamefont {A.~V.}\ \bibnamefont
  {Chumak}}, \bibinfo {author} {\bibfnamefont {V.}~\bibnamefont {Vasyuchka}},
  \bibinfo {author} {\bibfnamefont {A.}~\bibnamefont {Serga}}, \ and\ \bibinfo
  {author} {\bibfnamefont {B.}~\bibnamefont {Hillebrands}},\ }\bibfield
  {title} {\enquote {\bibinfo {title} {{Magnon spintronics}},}\ }\href
  {\doibase 10.1038/nphys3347} {\bibfield  {journal} {\bibinfo  {journal}
  {Nature Physics}\ }\textbf {\bibinfo {volume} {11}},\ \bibinfo {pages} {453}
  (\bibinfo {year} {2015})}\BibitemShut {NoStop}%
\bibitem [{\citenamefont {Jungwirth}\ \emph {et~al.}(2016)\citenamefont
  {Jungwirth}, \citenamefont {Marti}, \citenamefont {Wadley},\ and\
  \citenamefont {Wunderlich}}]{Jungwirth2016AntiferromagneticSpintronics}%
  \BibitemOpen
  \bibfield  {author} {\bibinfo {author} {\bibfnamefont {T.}~\bibnamefont
  {Jungwirth}}, \bibinfo {author} {\bibfnamefont {X.}~\bibnamefont {Marti}},
  \bibinfo {author} {\bibfnamefont {P.}~\bibnamefont {Wadley}}, \ and\ \bibinfo
  {author} {\bibfnamefont {J.}~\bibnamefont {Wunderlich}},\ }\bibfield  {title}
  {\enquote {\bibinfo {title} {{Antiferromagnetic spintronics}},}\ }\href
  {\doibase 10.1038/nnano.2016.18} {\bibfield  {journal} {\bibinfo  {journal}
  {Nature Nanotechnology}\ }\textbf {\bibinfo {volume} {11}},\ \bibinfo {pages}
  {231} (\bibinfo {year} {2016})}\BibitemShut {NoStop}%
\bibitem [{\citenamefont {Baltz}\ \emph {et~al.}(2018)\citenamefont {Baltz},
  \citenamefont {Manchon}, \citenamefont {Tsoi}, \citenamefont {Moriyama},
  \citenamefont {Ono},\ and\ \citenamefont
  {Tserkovnyak}}]{Baltz2018AntiferromagneticSpintronics}%
  \BibitemOpen
  \bibfield  {author} {\bibinfo {author} {\bibfnamefont {V.}~\bibnamefont
  {Baltz}}, \bibinfo {author} {\bibfnamefont {A.}~\bibnamefont {Manchon}},
  \bibinfo {author} {\bibfnamefont {M.}~\bibnamefont {Tsoi}}, \bibinfo {author}
  {\bibfnamefont {T.}~\bibnamefont {Moriyama}}, \bibinfo {author}
  {\bibfnamefont {T.}~\bibnamefont {Ono}}, \ and\ \bibinfo {author}
  {\bibfnamefont {Y.}~\bibnamefont {Tserkovnyak}},\ }\bibfield  {title}
  {\enquote {\bibinfo {title} {{Antiferromagnetic spintronics}},}\ }\href
  {\doibase 10.1103/RevModPhys.90.015005} {\bibfield  {journal} {\bibinfo
  {journal} {Reviews of Modern Physics}\ }\textbf {\bibinfo {volume} {90}},\
  \bibinfo {pages} {015005} (\bibinfo {year} {2018})}\BibitemShut {NoStop}%
\bibitem [{\citenamefont {Xiao}\ \emph {et~al.}(2010)\citenamefont {Xiao},
  \citenamefont {Bauer}, \citenamefont {Uchida}, \citenamefont {Saitoh},\ and\
  \citenamefont {Maekawa}}]{Xiao2010TheoryEffect}%
  \BibitemOpen
  \bibfield  {author} {\bibinfo {author} {\bibfnamefont {J.}~\bibnamefont
  {Xiao}}, \bibinfo {author} {\bibfnamefont {G.~E.~W.}\ \bibnamefont {Bauer}},
  \bibinfo {author} {\bibfnamefont {K.-c.}\ \bibnamefont {Uchida}}, \bibinfo
  {author} {\bibfnamefont {E.}~\bibnamefont {Saitoh}}, \ and\ \bibinfo {author}
  {\bibfnamefont {S.}~\bibnamefont {Maekawa}},\ }\bibfield  {title} {\enquote
  {\bibinfo {title} {{Theory of magnon-driven spin Seebeck effect}},}\ }\href
  {\doibase 10.1103/PhysRevB.81.214418} {\bibfield  {journal} {\bibinfo
  {journal} {Physical Review B}\ }\textbf {\bibinfo {volume} {81}},\ \bibinfo
  {pages} {214418} (\bibinfo {year} {2010})}\BibitemShut {NoStop}%
\bibitem [{\citenamefont {Zyuzin}\ and\ \citenamefont
  {Kovalev}(2016)}]{Zyuzin2016MagnonAntiferromagnets}%
  \BibitemOpen
  \bibfield  {author} {\bibinfo {author} {\bibfnamefont {V.~A.}\ \bibnamefont
  {Zyuzin}}\ and\ \bibinfo {author} {\bibfnamefont {A.~A.}\ \bibnamefont
  {Kovalev}},\ }\bibfield  {title} {\enquote {\bibinfo {title} {{Magnon Spin
  Nernst Effect in Antiferromagnets}},}\ }\href {\doibase
  10.1103/PhysRevLett.117.217203} {\bibfield  {journal} {\bibinfo  {journal}
  {Physical Review Letters}\ }\textbf {\bibinfo {volume} {117}},\ \bibinfo
  {pages} {217203} (\bibinfo {year} {2016})}\BibitemShut {NoStop}%
\bibitem [{\citenamefont {Cheng}\ \emph {et~al.}(2016)\citenamefont {Cheng},
  \citenamefont {Okamoto},\ and\ \citenamefont
  {Xiao}}]{Cheng2016SpinAntiferromagnets}%
  \BibitemOpen
  \bibfield  {author} {\bibinfo {author} {\bibfnamefont {R.}~\bibnamefont
  {Cheng}}, \bibinfo {author} {\bibfnamefont {S.}~\bibnamefont {Okamoto}}, \
  and\ \bibinfo {author} {\bibfnamefont {D.}~\bibnamefont {Xiao}},\ }\bibfield
  {title} {\enquote {\bibinfo {title} {{Spin Nernst Effect of Magnons in
  Collinear Antiferromagnets}},}\ }\href {\doibase
  10.1103/PhysRevLett.117.217202} {\bibfield  {journal} {\bibinfo  {journal}
  {Physical Review Letters}\ }\textbf {\bibinfo {volume} {117}},\ \bibinfo
  {pages} {217202} (\bibinfo {year} {2016})}\BibitemShut {NoStop}%
\bibitem [{\citenamefont {Shiomi}\ \emph {et~al.}(2017)\citenamefont {Shiomi},
  \citenamefont {Takashima},\ and\ \citenamefont
  {Saitoh}}]{Shiomi2017ExperimentalMnPS3}%
  \BibitemOpen
  \bibfield  {author} {\bibinfo {author} {\bibfnamefont {Y.}~\bibnamefont
  {Shiomi}}, \bibinfo {author} {\bibfnamefont {R.}~\bibnamefont {Takashima}}, \
  and\ \bibinfo {author} {\bibfnamefont {E.}~\bibnamefont {Saitoh}},\
  }\bibfield  {title} {\enquote {\bibinfo {title} {{Experimental evidence
  consistent with a magnon Nernst effect in the antiferromagnetic insulator
  MnPS3}},}\ }\href {\doibase 10.1103/PhysRevB.96.134425} {\bibfield  {journal}
  {\bibinfo  {journal} {Physical Review B}\ }\textbf {\bibinfo {volume} {96}},\
  \bibinfo {pages} {134425} (\bibinfo {year} {2017})}\BibitemShut {NoStop}%
\bibitem [{\citenamefont {Katsura}\ \emph {et~al.}(2010)\citenamefont
  {Katsura}, \citenamefont {Nagaosa},\ and\ \citenamefont
  {Lee}}]{Katsura2010TheoryMagnets}%
  \BibitemOpen
  \bibfield  {author} {\bibinfo {author} {\bibfnamefont {H.}~\bibnamefont
  {Katsura}}, \bibinfo {author} {\bibfnamefont {N.}~\bibnamefont {Nagaosa}}, \
  and\ \bibinfo {author} {\bibfnamefont {P.~A.}\ \bibnamefont {Lee}},\
  }\bibfield  {title} {\enquote {\bibinfo {title} {{Theory of the Thermal Hall
  Effect in Quantum Magnets}},}\ }\href {\doibase
  10.1103/PhysRevLett.104.066403} {\bibfield  {journal} {\bibinfo  {journal}
  {Physical Review Letters}\ }\textbf {\bibinfo {volume} {104}},\ \bibinfo
  {pages} {066403} (\bibinfo {year} {2010})}\BibitemShut {NoStop}%
\bibitem [{\citenamefont {Onose}\ \emph {et~al.}(2010)\citenamefont {Onose},
  \citenamefont {Ideue}, \citenamefont {Katsura}, \citenamefont {Shiomi},
  \citenamefont {Nagaosa},\ and\ \citenamefont
  {Tokura}}]{Onose2010ObservationEffect}%
  \BibitemOpen
  \bibfield  {author} {\bibinfo {author} {\bibfnamefont {Y.}~\bibnamefont
  {Onose}}, \bibinfo {author} {\bibfnamefont {T.}~\bibnamefont {Ideue}},
  \bibinfo {author} {\bibfnamefont {H.}~\bibnamefont {Katsura}}, \bibinfo
  {author} {\bibfnamefont {Y.}~\bibnamefont {Shiomi}}, \bibinfo {author}
  {\bibfnamefont {N.}~\bibnamefont {Nagaosa}}, \ and\ \bibinfo {author}
  {\bibfnamefont {Y.}~\bibnamefont {Tokura}},\ }\bibfield  {title} {\enquote
  {\bibinfo {title} {{Observation of the Magnon Hall Effect}},}\ }\href
  {\doibase 10.1126/science.1188260} {\bibfield  {journal} {\bibinfo  {journal}
  {Science}\ }\textbf {\bibinfo {volume} {329}},\ \bibinfo {pages} {297}
  (\bibinfo {year} {2010})}\BibitemShut {NoStop}%
\bibitem [{\citenamefont {Matsumoto}\ and\ \citenamefont
  {Murakami}(2011{\natexlab{a}})}]{Matsumoto2011RotationalEffect}%
  \BibitemOpen
  \bibfield  {author} {\bibinfo {author} {\bibfnamefont {R.}~\bibnamefont
  {Matsumoto}}\ and\ \bibinfo {author} {\bibfnamefont {S.}~\bibnamefont
  {Murakami}},\ }\bibfield  {title} {\enquote {\bibinfo {title} {{Rotational
  motion of magnons and the thermal Hall effect}},}\ }\href {\doibase
  10.1103/PhysRevB.84.184406} {\bibfield  {journal} {\bibinfo  {journal}
  {Physical Review B}\ }\textbf {\bibinfo {volume} {84}},\ \bibinfo {pages}
  {184406} (\bibinfo {year} {2011}{\natexlab{a}})}\BibitemShut {NoStop}%
\bibitem [{\citenamefont {Matsumoto}\ and\ \citenamefont
  {Murakami}(2011{\natexlab{b}})}]{Matsumoto2011TheoreticalFerromagnets}%
  \BibitemOpen
  \bibfield  {author} {\bibinfo {author} {\bibfnamefont {R.}~\bibnamefont
  {Matsumoto}}\ and\ \bibinfo {author} {\bibfnamefont {S.}~\bibnamefont
  {Murakami}},\ }\bibfield  {title} {\enquote {\bibinfo {title} {{Theoretical
  Prediction of a Rotating Magnon Wave Packet in Ferromagnets}},}\ }\href
  {\doibase 10.1103/PhysRevLett.106.197202} {\bibfield  {journal} {\bibinfo
  {journal} {Physical Review Letters}\ }\textbf {\bibinfo {volume} {106}},\
  \bibinfo {pages} {197202} (\bibinfo {year} {2011}{\natexlab{b}})}\BibitemShut
  {NoStop}%
\bibitem [{\citenamefont {Matsumoto}\ \emph {et~al.}(2014)\citenamefont
  {Matsumoto}, \citenamefont {Shindou},\ and\ \citenamefont
  {Murakami}}]{Matsumoto2014ThermalInteraction}%
  \BibitemOpen
  \bibfield  {author} {\bibinfo {author} {\bibfnamefont {R.}~\bibnamefont
  {Matsumoto}}, \bibinfo {author} {\bibfnamefont {R.}~\bibnamefont {Shindou}},
  \ and\ \bibinfo {author} {\bibfnamefont {S.}~\bibnamefont {Murakami}},\
  }\bibfield  {title} {\enquote {\bibinfo {title} {{Thermal Hall effect of
  magnons in magnets with dipolar interaction}},}\ }\href {\doibase
  10.1103/PhysRevB.89.054420} {\bibfield  {journal} {\bibinfo  {journal}
  {Physical Review B}\ }\textbf {\bibinfo {volume} {89}},\ \bibinfo {pages}
  {054420} (\bibinfo {year} {2014})}\BibitemShut {NoStop}%
\bibitem [{\citenamefont {Zhang}\ \emph {et~al.}(2018)\citenamefont {Zhang},
  \citenamefont {Okamoto},\ and\ \citenamefont
  {Xiao}}]{Zhang2018Spin-NernstInsulator}%
  \BibitemOpen
  \bibfield  {author} {\bibinfo {author} {\bibfnamefont {Y.}~\bibnamefont
  {Zhang}}, \bibinfo {author} {\bibfnamefont {S.}~\bibnamefont {Okamoto}}, \
  and\ \bibinfo {author} {\bibfnamefont {D.}~\bibnamefont {Xiao}},\ }\bibfield
  {title} {\enquote {\bibinfo {title} {{Spin-Nernst effect in the paramagnetic
  regime of an antiferromagnetic insulator}},}\ }\href {\doibase
  10.1103/PhysRevB.98.035424} {\bibfield  {journal} {\bibinfo  {journal}
  {Physical Review B}\ }\textbf {\bibinfo {volume} {98}},\ \bibinfo {pages}
  {035424} (\bibinfo {year} {2018})}\BibitemShut {NoStop}%
\bibitem [{\citenamefont {Park}\ and\ \citenamefont
  {Yang}(2019)}]{Park2019TopologicalAntiferromagnets}%
  \BibitemOpen
  \bibfield  {author} {\bibinfo {author} {\bibfnamefont {S.}~\bibnamefont
  {Park}}\ and\ \bibinfo {author} {\bibfnamefont {B.-J.}\ \bibnamefont
  {Yang}},\ }\bibfield  {title} {\enquote {\bibinfo {title} {{Topological
  magnetoelastic excitations in noncollinear antiferromagnets}},}\ }\href
  {\doibase 10.1103/PhysRevB.99.174435} {\bibfield  {journal} {\bibinfo
  {journal} {Physical Review B}\ }\textbf {\bibinfo {volume} {99}},\ \bibinfo
  {pages} {174435} (\bibinfo {year} {2019})}\BibitemShut {NoStop}%
\bibitem [{\citenamefont {Laurell}\ and\ \citenamefont
  {Fiete}(2018)}]{Laurell2018MagnonInteractions}%
  \BibitemOpen
  \bibfield  {author} {\bibinfo {author} {\bibfnamefont {P.}~\bibnamefont
  {Laurell}}\ and\ \bibinfo {author} {\bibfnamefont {G.~A.}\ \bibnamefont
  {Fiete}},\ }\bibfield  {title} {\enquote {\bibinfo {title} {{Magnon thermal
  Hall effect in kagome antiferromagnets with Dzyaloshinskii-Moriya
  interactions}},}\ }\href {\doibase 10.1103/PhysRevB.98.094419} {\bibfield
  {journal} {\bibinfo  {journal} {Physical Review B}\ }\textbf {\bibinfo
  {volume} {98}},\ \bibinfo {pages} {094419} (\bibinfo {year}
  {2018})}\BibitemShut {NoStop}%
\bibitem [{\citenamefont {Doki}\ \emph {et~al.}(2018)\citenamefont {Doki},
  \citenamefont {Akazawa}, \citenamefont {Lee}, \citenamefont {Han},
  \citenamefont {Sugii}, \citenamefont {Shimozawa}, \citenamefont {Kawashima},
  \citenamefont {Oda}, \citenamefont {Yoshida},\ and\ \citenamefont
  {Yamashita}}]{Doki2018SpinAntiferromagnet}%
  \BibitemOpen
  \bibfield  {author} {\bibinfo {author} {\bibfnamefont {H.}~\bibnamefont
  {Doki}}, \bibinfo {author} {\bibfnamefont {M.}~\bibnamefont {Akazawa}},
  \bibinfo {author} {\bibfnamefont {H.-Y.}\ \bibnamefont {Lee}}, \bibinfo
  {author} {\bibfnamefont {J.~H.}\ \bibnamefont {Han}}, \bibinfo {author}
  {\bibfnamefont {K.}~\bibnamefont {Sugii}}, \bibinfo {author} {\bibfnamefont
  {M.}~\bibnamefont {Shimozawa}}, \bibinfo {author} {\bibfnamefont
  {N.}~\bibnamefont {Kawashima}}, \bibinfo {author} {\bibfnamefont
  {M.}~\bibnamefont {Oda}}, \bibinfo {author} {\bibfnamefont {H.}~\bibnamefont
  {Yoshida}}, \ and\ \bibinfo {author} {\bibfnamefont {M.}~\bibnamefont
  {Yamashita}},\ }\bibfield  {title} {\enquote {\bibinfo {title} {{Spin Thermal
  Hall Conductivity of a Kagome Antiferromagnet}},}\ }\href {\doibase
  10.1103/PhysRevLett.121.097203} {\bibfield  {journal} {\bibinfo  {journal}
  {Physical Review Letters}\ }\textbf {\bibinfo {volume} {121}},\ \bibinfo
  {pages} {097203} (\bibinfo {year} {2018})}\BibitemShut {NoStop}%
\bibitem [{\citenamefont {Mook}\ \emph {et~al.}(2019)\citenamefont {Mook},
  \citenamefont {Henk},\ and\ \citenamefont
  {Mertig}}]{Mook2019ThermalAntiferromagnets}%
  \BibitemOpen
  \bibfield  {author} {\bibinfo {author} {\bibfnamefont {A.}~\bibnamefont
  {Mook}}, \bibinfo {author} {\bibfnamefont {J.}~\bibnamefont {Henk}}, \ and\
  \bibinfo {author} {\bibfnamefont {I.}~\bibnamefont {Mertig}},\ }\bibfield
  {title} {\enquote {\bibinfo {title} {{Thermal Hall effect in noncollinear
  coplanar insulating antiferromagnets}},}\ }\href {\doibase
  10.1103/PhysRevB.99.014427} {\bibfield  {journal} {\bibinfo  {journal}
  {Physical Review B}\ }\textbf {\bibinfo {volume} {99}} (\bibinfo {year}
  {2019}),\ 10.1103/PhysRevB.99.014427}\BibitemShut {NoStop}%
\bibitem [{\citenamefont
  {Owerre}(2017{\natexlab{a}})}]{Owerre2017TopologicalAntiferromagnets}%
  \BibitemOpen
  \bibfield  {author} {\bibinfo {author} {\bibfnamefont {S.~A.}\ \bibnamefont
  {Owerre}},\ }\bibfield  {title} {\enquote {\bibinfo {title} {{Topological
  thermal Hall effect in frustrated kagome antiferromagnets}},}\ }\href
  {\doibase 10.1103/PhysRevB.95.014422} {\bibfield  {journal} {\bibinfo
  {journal} {Physical Review B}\ }\textbf {\bibinfo {volume} {95}},\ \bibinfo
  {pages} {014422} (\bibinfo {year} {2017}{\natexlab{a}})}\BibitemShut
  {NoStop}%
\bibitem [{\citenamefont
  {Owerre}(2017{\natexlab{b}})}]{Owerre2017NoncollinearInsulator}%
  \BibitemOpen
  \bibfield  {author} {\bibinfo {author} {\bibfnamefont {S.~A.}\ \bibnamefont
  {Owerre}},\ }\bibfield  {title} {\enquote {\bibinfo {title} {{Noncollinear
  antiferromagnetic Haldane magnon insulator}},}\ }\href {\doibase
  10.1063/1.4985615} {\bibfield  {journal} {\bibinfo  {journal} {Journal of
  Applied Physics}\ }\textbf {\bibinfo {volume} {121}},\ \bibinfo {pages}
  {223904} (\bibinfo {year} {2017}{\natexlab{b}})}\BibitemShut {NoStop}%
\bibitem [{\citenamefont {Ideue}\ \emph {et~al.}(2012)\citenamefont {Ideue},
  \citenamefont {Onose}, \citenamefont {Katsura}, \citenamefont {Shiomi},
  \citenamefont {Ishiwata}, \citenamefont {Nagaosa},\ and\ \citenamefont
  {Tokura}}]{Ideue2012EffectInsulators}%
  \BibitemOpen
  \bibfield  {author} {\bibinfo {author} {\bibfnamefont {T.}~\bibnamefont
  {Ideue}}, \bibinfo {author} {\bibfnamefont {Y.}~\bibnamefont {Onose}},
  \bibinfo {author} {\bibfnamefont {H.}~\bibnamefont {Katsura}}, \bibinfo
  {author} {\bibfnamefont {Y.}~\bibnamefont {Shiomi}}, \bibinfo {author}
  {\bibfnamefont {S.}~\bibnamefont {Ishiwata}}, \bibinfo {author}
  {\bibfnamefont {N.}~\bibnamefont {Nagaosa}}, \ and\ \bibinfo {author}
  {\bibfnamefont {Y.}~\bibnamefont {Tokura}},\ }\bibfield  {title} {\enquote
  {\bibinfo {title} {{Effect of lattice geometry on magnon Hall effect in
  ferromagnetic insulators}},}\ }\href {\doibase 10.1103/PhysRevB.85.134411}
  {\bibfield  {journal} {\bibinfo  {journal} {Physical Review B}\ }\textbf
  {\bibinfo {volume} {85}},\ \bibinfo {pages} {134411} (\bibinfo {year}
  {2012})}\BibitemShut {NoStop}%
\bibitem [{\citenamefont {Kawano}\ and\ \citenamefont
  {Hotta}(2019)}]{Kawano2019ThermalAntiferromagnet}%
  \BibitemOpen
  \bibfield  {author} {\bibinfo {author} {\bibfnamefont {M.}~\bibnamefont
  {Kawano}}\ and\ \bibinfo {author} {\bibfnamefont {C.}~\bibnamefont {Hotta}},\
  }\bibfield  {title} {\enquote {\bibinfo {title} {{Thermal Hall effect and
  topological edge states in a square-lattice antiferromagnet}},}\ }\href
  {\doibase 10.1103/PhysRevB.99.054422} {\bibfield  {journal} {\bibinfo
  {journal} {Physical Review B}\ }\textbf {\bibinfo {volume} {99}},\ \bibinfo
  {pages} {054422} (\bibinfo {year} {2019})}\BibitemShut {NoStop}%
\bibitem [{\citenamefont
  {Dzyaloshinsky}(1958)}]{Dzyaloshinsky1958AAntiferromagnetics}%
  \BibitemOpen
  \bibfield  {author} {\bibinfo {author} {\bibfnamefont {I.}~\bibnamefont
  {Dzyaloshinsky}},\ }\bibfield  {title} {\enquote {\bibinfo {title} {{A
  thermodynamic theory of “weak” ferromagnetism of antiferromagnetics}},}\
  }\href {\doibase 10.1016/0022-3697(58)90076-3} {\bibfield  {journal}
  {\bibinfo  {journal} {Journal of Physics and Chemistry of Solids}\ }\textbf
  {\bibinfo {volume} {4}},\ \bibinfo {pages} {241} (\bibinfo {year}
  {1958})}\BibitemShut {NoStop}%
\bibitem [{\citenamefont {Moriya}(1960)}]{Moriya1960AnisotropicFerromagnetism}%
  \BibitemOpen
  \bibfield  {author} {\bibinfo {author} {\bibfnamefont {T.}~\bibnamefont
  {Moriya}},\ }\bibfield  {title} {\enquote {\bibinfo {title} {{Anisotropic
  Superexchange Interaction and Weak Ferromagnetism}},}\ }\href {\doibase
  10.1103/PhysRev.120.91} {\bibfield  {journal} {\bibinfo  {journal} {Physical
  Review}\ }\textbf {\bibinfo {volume} {120}},\ \bibinfo {pages} {91} (\bibinfo
  {year} {1960})}\BibitemShut {NoStop}%
\bibitem [{\citenamefont
  {Owerre}(2017{\natexlab{c}})}]{Owerre2017TopologicalLattice}%
  \BibitemOpen
  \bibfield  {author} {\bibinfo {author} {\bibfnamefont {S.~A.}\ \bibnamefont
  {Owerre}},\ }\bibfield  {title} {\enquote {\bibinfo {title} {{Topological
  magnon bands and unconventional thermal Hall effect on the frustrated
  honeycomb and bilayer triangular lattice}},}\ }\href {\doibase
  10.1088/1361-648X/aa7dd2} {\bibfield  {journal} {\bibinfo  {journal} {Journal
  of Physics: Condensed Matter}\ }\textbf {\bibinfo {volume} {29}},\ \bibinfo
  {pages} {385801} (\bibinfo {year} {2017}{\natexlab{c}})}\BibitemShut
  {NoStop}%
\bibitem [{\citenamefont {G{\'{o}}mez~Albarrac{\'{i}}n}\ \emph
  {et~al.}(2021)\citenamefont {G{\'{o}}mez~Albarrac{\'{i}}n}, \citenamefont
  {Rosales},\ and\ \citenamefont {Pujol}}]{GomezAlbarracin2021ChiralLattice}%
  \BibitemOpen
  \bibfield  {author} {\bibinfo {author} {\bibfnamefont {F.~A.}\ \bibnamefont
  {G{\'{o}}mez~Albarrac{\'{i}}n}}, \bibinfo {author} {\bibfnamefont {H.~D.}\
  \bibnamefont {Rosales}}, \ and\ \bibinfo {author} {\bibfnamefont
  {P.}~\bibnamefont {Pujol}},\ }\bibfield  {title} {\enquote {\bibinfo {title}
  {{Chiral phase transition and thermal Hall effect in an anisotropic spin
  model on the kagome lattice}},}\ }\href {\doibase
  10.1103/PhysRevB.103.054405} {\bibfield  {journal} {\bibinfo  {journal}
  {Physical Review B}\ }\textbf {\bibinfo {volume} {103}},\ \bibinfo {pages}
  {054405} (\bibinfo {year} {2021})}\BibitemShut {NoStop}%
\bibitem [{\citenamefont
  {Owerre}(2017{\natexlab{d}})}]{Owerre2017MagnonInteraction}%
  \BibitemOpen
  \bibfield  {author} {\bibinfo {author} {\bibfnamefont {S.~A.}\ \bibnamefont
  {Owerre}},\ }\bibfield  {title} {\enquote {\bibinfo {title} {{Magnon Hall
  effect without Dzyaloshinskii–Moriya interaction}},}\ }\href {\doibase
  10.1088/0953-8984/29/3/03LT01} {\bibfield  {journal} {\bibinfo  {journal}
  {Journal of Physics: Condensed Matter}\ }\textbf {\bibinfo {volume} {29}},\
  \bibinfo {pages} {03LT01} (\bibinfo {year} {2017}{\natexlab{d}})}\BibitemShut
  {NoStop}%
\bibitem [{\citenamefont {Naka}\ \emph {et~al.}(2019)\citenamefont {Naka},
  \citenamefont {Hayami}, \citenamefont {Kusunose}, \citenamefont {Yanagi},
  \citenamefont {Motome},\ and\ \citenamefont
  {Seo}}]{Naka2019SpinAntiferromagnets}%
  \BibitemOpen
  \bibfield  {author} {\bibinfo {author} {\bibfnamefont {M.}~\bibnamefont
  {Naka}}, \bibinfo {author} {\bibfnamefont {S.}~\bibnamefont {Hayami}},
  \bibinfo {author} {\bibfnamefont {H.}~\bibnamefont {Kusunose}}, \bibinfo
  {author} {\bibfnamefont {Y.}~\bibnamefont {Yanagi}}, \bibinfo {author}
  {\bibfnamefont {Y.}~\bibnamefont {Motome}}, \ and\ \bibinfo {author}
  {\bibfnamefont {H.}~\bibnamefont {Seo}},\ }\bibfield  {title} {\enquote
  {\bibinfo {title} {{Spin current generation in organic antiferromagnets}},}\
  }\href {\doibase 10.1038/s41467-019-12229-y} {\bibfield  {journal} {\bibinfo
  {journal} {Nature Communications}\ }\textbf {\bibinfo {volume} {10}},\
  \bibinfo {pages} {4305} (\bibinfo {year} {2019})}\BibitemShut {NoStop}%
\bibitem [{\citenamefont {Naka}\ \emph {et~al.}(2020)\citenamefont {Naka},
  \citenamefont {Hayami}, \citenamefont {Kusunose}, \citenamefont {Yanagi},
  \citenamefont {Motome},\ and\ \citenamefont
  {Seo}}]{Naka2020AnomalousAntiferromagnets}%
  \BibitemOpen
  \bibfield  {author} {\bibinfo {author} {\bibfnamefont {M.}~\bibnamefont
  {Naka}}, \bibinfo {author} {\bibfnamefont {S.}~\bibnamefont {Hayami}},
  \bibinfo {author} {\bibfnamefont {H.}~\bibnamefont {Kusunose}}, \bibinfo
  {author} {\bibfnamefont {Y.}~\bibnamefont {Yanagi}}, \bibinfo {author}
  {\bibfnamefont {Y.}~\bibnamefont {Motome}}, \ and\ \bibinfo {author}
  {\bibfnamefont {H.}~\bibnamefont {Seo}},\ }\bibfield  {title} {\enquote
  {\bibinfo {title} {{Anomalous Hall effect in {$\kappa$}-type organic
  antiferromagnets}},}\ }\href {\doibase 10.1103/PhysRevB.102.075112}
  {\bibfield  {journal} {\bibinfo  {journal} {Physical Review B}\ }\textbf
  {\bibinfo {volume} {102}},\ \bibinfo {pages} {075112} (\bibinfo {year}
  {2020})}\BibitemShut {NoStop}%
\bibitem [{\citenamefont {Naka}\ \emph {et~al.}(2021)\citenamefont {Naka},
  \citenamefont {Motome},\ and\ \citenamefont
  {Seo}}]{Naka2021PerovskiteGenerator}%
  \BibitemOpen
  \bibfield  {author} {\bibinfo {author} {\bibfnamefont {M.}~\bibnamefont
  {Naka}}, \bibinfo {author} {\bibfnamefont {Y.}~\bibnamefont {Motome}}, \ and\
  \bibinfo {author} {\bibfnamefont {H.}~\bibnamefont {Seo}},\ }\bibfield
  {title} {\enquote {\bibinfo {title} {{Perovskite as a spin current
  generator}},}\ }\href {\doibase 10.1103/PhysRevB.103.125114} {\bibfield
  {journal} {\bibinfo  {journal} {Physical Review B}\ }\textbf {\bibinfo
  {volume} {103}},\ \bibinfo {pages} {125114} (\bibinfo {year}
  {2021})}\BibitemShut {NoStop}%
\bibitem [{\citenamefont {Lacroix}\ \emph {et~al.}(2011)\citenamefont
  {Lacroix}, \citenamefont {Mendels},\ and\ \citenamefont
  {Mila}}]{2011IntroductionMagnetism}%
  \BibitemOpen
  \bibinfo {editor} {\bibfnamefont {C.}~\bibnamefont {Lacroix}}, \bibinfo
  {editor} {\bibfnamefont {P.}~\bibnamefont {Mendels}}, \ and\ \bibinfo
  {editor} {\bibfnamefont {F.}~\bibnamefont {Mila}},\ eds.,\ \href {\doibase
  10.1007/978-3-642-10589-0} {\emph {\bibinfo {title} {{Introduction to
  Frustrated Magnetism}}}},\ Vol.\ \bibinfo {volume} {164}\ (\bibinfo
  {publisher} {Springer Berlin Heidelberg},\ \bibinfo {address} {Berlin,
  Heidelberg},\ \bibinfo {year} {2011})\BibitemShut {NoStop}%
\bibitem [{\citenamefont {{\v{Z}}elezn{\'{y}}}\ \emph
  {et~al.}(2017)\citenamefont {{\v{Z}}elezn{\'{y}}}, \citenamefont {Zhang},
  \citenamefont {Felser},\ and\ \citenamefont
  {Yan}}]{Zelezny2017Spin-PolarizedAntiferromagnets}%
  \BibitemOpen
  \bibfield  {author} {\bibinfo {author} {\bibfnamefont {J.}~\bibnamefont
  {{\v{Z}}elezn{\'{y}}}}, \bibinfo {author} {\bibfnamefont {Y.}~\bibnamefont
  {Zhang}}, \bibinfo {author} {\bibfnamefont {C.}~\bibnamefont {Felser}}, \
  and\ \bibinfo {author} {\bibfnamefont {B.}~\bibnamefont {Yan}},\ }\bibfield
  {title} {\enquote {\bibinfo {title} {{Spin-Polarized Current in Noncollinear
  Antiferromagnets}},}\ }\href {\doibase 10.1103/PhysRevLett.119.187204}
  {\bibfield  {journal} {\bibinfo  {journal} {Physical Review Letters}\
  }\textbf {\bibinfo {volume} {119}},\ \bibinfo {pages} {187204} (\bibinfo
  {year} {2017})}\BibitemShut {NoStop}%
\bibitem [{\citenamefont {Kimata}\ \emph {et~al.}(2019)\citenamefont {Kimata},
  \citenamefont {Chen}, \citenamefont {Kondou}, \citenamefont {Sugimoto},
  \citenamefont {Muduli}, \citenamefont {Ikhlas}, \citenamefont {Omori},
  \citenamefont {Tomita}, \citenamefont {MacDonald}, \citenamefont
  {Nakatsuji},\ and\ \citenamefont
  {Otani}}]{Kimata2019MagneticAntiferromagnet}%
  \BibitemOpen
  \bibfield  {author} {\bibinfo {author} {\bibfnamefont {M.}~\bibnamefont
  {Kimata}}, \bibinfo {author} {\bibfnamefont {H.}~\bibnamefont {Chen}},
  \bibinfo {author} {\bibfnamefont {K.}~\bibnamefont {Kondou}}, \bibinfo
  {author} {\bibfnamefont {S.}~\bibnamefont {Sugimoto}}, \bibinfo {author}
  {\bibfnamefont {P.~K.}\ \bibnamefont {Muduli}}, \bibinfo {author}
  {\bibfnamefont {M.}~\bibnamefont {Ikhlas}}, \bibinfo {author} {\bibfnamefont
  {Y.}~\bibnamefont {Omori}}, \bibinfo {author} {\bibfnamefont
  {T.}~\bibnamefont {Tomita}}, \bibinfo {author} {\bibfnamefont {A.~H.}\
  \bibnamefont {MacDonald}}, \bibinfo {author} {\bibfnamefont {S.}~\bibnamefont
  {Nakatsuji}}, \ and\ \bibinfo {author} {\bibfnamefont {Y.}~\bibnamefont
  {Otani}},\ }\bibfield  {title} {\enquote {\bibinfo {title} {{Magnetic and
  magnetic inverse spin Hall effects in a non-collinear antiferromagnet}},}\
  }\href {\doibase 10.1038/s41586-018-0853-0} {\bibfield  {journal} {\bibinfo
  {journal} {Nature}\ }\textbf {\bibinfo {volume} {565}},\ \bibinfo {pages}
  {627} (\bibinfo {year} {2019})}\BibitemShut {NoStop}%
\bibitem [{\citenamefont {Li}\ \emph {et~al.}(2020{\natexlab{a}})\citenamefont
  {Li}, \citenamefont {Sandhoefner},\ and\ \citenamefont
  {Kovalev}}]{Li2020IntrinsicAntiferromagnet}%
  \BibitemOpen
  \bibfield  {author} {\bibinfo {author} {\bibfnamefont {B.}~\bibnamefont
  {Li}}, \bibinfo {author} {\bibfnamefont {S.}~\bibnamefont {Sandhoefner}}, \
  and\ \bibinfo {author} {\bibfnamefont {A.~A.}\ \bibnamefont {Kovalev}},\
  }\bibfield  {title} {\enquote {\bibinfo {title} {{Intrinsic spin Nernst
  effect of magnons in a noncollinear antiferromagnet}},}\ }\href {\doibase
  10.1103/PhysRevResearch.2.013079} {\bibfield  {journal} {\bibinfo  {journal}
  {Physical Review Research}\ }\textbf {\bibinfo {volume} {2}},\ \bibinfo
  {pages} {013079} (\bibinfo {year} {2020}{\natexlab{a}})}\BibitemShut
  {NoStop}%
\bibitem [{\citenamefont {Park}\ \emph {et~al.}(2020)\citenamefont {Park},
  \citenamefont {Nagaosa},\ and\ \citenamefont
  {Yang}}]{Park2020ThermalInteraction}%
  \BibitemOpen
  \bibfield  {author} {\bibinfo {author} {\bibfnamefont {S.}~\bibnamefont
  {Park}}, \bibinfo {author} {\bibfnamefont {N.}~\bibnamefont {Nagaosa}}, \
  and\ \bibinfo {author} {\bibfnamefont {B.-J.}\ \bibnamefont {Yang}},\
  }\bibfield  {title} {\enquote {\bibinfo {title} {{Thermal Hall Effect, Spin
  Nernst Effect, and Spin Density Induced by a Thermal Gradient in Collinear
  Ferrimagnets from Magnon–Phonon Interaction}},}\ }\href {\doibase
  10.1021/acs.nanolett.0c00363} {\bibfield  {journal} {\bibinfo  {journal}
  {Nano Letters}\ }\textbf {\bibinfo {volume} {20}},\ \bibinfo {pages} {2741}
  (\bibinfo {year} {2020})}\BibitemShut {NoStop}%
\bibitem [{\citenamefont {Holstein}\ and\ \citenamefont
  {Primakoff}(1940)}]{Holstein1940FieldFerromagnet}%
  \BibitemOpen
  \bibfield  {author} {\bibinfo {author} {\bibfnamefont {T.}~\bibnamefont
  {Holstein}}\ and\ \bibinfo {author} {\bibfnamefont {H.}~\bibnamefont
  {Primakoff}},\ }\bibfield  {title} {\enquote {\bibinfo {title} {{Field
  Dependence of the Intrinsic Domain Magnetization of a Ferromagnet}},}\ }\href
  {\doibase 10.1103/PhysRev.58.1098} {\bibfield  {journal} {\bibinfo  {journal}
  {Physical Review}\ }\textbf {\bibinfo {volume} {58}},\ \bibinfo {pages}
  {1098} (\bibinfo {year} {1940})}\BibitemShut {NoStop}%
\bibitem [{\citenamefont {Zhitomirsky}\ and\ \citenamefont
  {Nikuni}(1998)}]{Zhitomirsky1998MagnetizationAntiferromagnet}%
  \BibitemOpen
  \bibfield  {author} {\bibinfo {author} {\bibfnamefont {M.~E.}\ \bibnamefont
  {Zhitomirsky}}\ and\ \bibinfo {author} {\bibfnamefont {T.}~\bibnamefont
  {Nikuni}},\ }\bibfield  {title} {\enquote {\bibinfo {title} {{Magnetization
  curve of a square-lattice Heisenberg antiferromagnet}},}\ }\href {\doibase
  10.1103/PhysRevB.57.5013} {\bibfield  {journal} {\bibinfo  {journal}
  {Physical Review B}\ }\textbf {\bibinfo {volume} {57}},\ \bibinfo {pages}
  {5013} (\bibinfo {year} {1998})}\BibitemShut {NoStop}%
\bibitem [{\citenamefont {Rastelli}\ \emph {et~al.}(1979)\citenamefont
  {Rastelli}, \citenamefont {Tassi},\ and\ \citenamefont
  {Reatto}}]{Rastelli1979Non-simpleHamiltonians}%
  \BibitemOpen
  \bibfield  {author} {\bibinfo {author} {\bibfnamefont {E.}~\bibnamefont
  {Rastelli}}, \bibinfo {author} {\bibfnamefont {A.}~\bibnamefont {Tassi}}, \
  and\ \bibinfo {author} {\bibfnamefont {L.}~\bibnamefont {Reatto}},\
  }\bibfield  {title} {\enquote {\bibinfo {title} {{Non-simple magnetic order
  for simple Hamiltonians}},}\ }\href {\doibase 10.1016/0378-4363(79)90002-0}
  {\bibfield  {journal} {\bibinfo  {journal} {Physica B+C}\ }\textbf {\bibinfo
  {volume} {97}},\ \bibinfo {pages} {1} (\bibinfo {year} {1979})}\BibitemShut
  {NoStop}%
\bibitem [{\citenamefont {Fouet}\ \emph {et~al.}(2001)\citenamefont {Fouet},
  \citenamefont {Sindzingre},\ and\ \citenamefont
  {Lhuillier}}]{Fouet2001AnLattice}%
  \BibitemOpen
  \bibfield  {author} {\bibinfo {author} {\bibfnamefont {J.}~\bibnamefont
  {Fouet}}, \bibinfo {author} {\bibfnamefont {P.}~\bibnamefont {Sindzingre}}, \
  and\ \bibinfo {author} {\bibfnamefont {C.}~\bibnamefont {Lhuillier}},\
  }\bibfield  {title} {\enquote {\bibinfo {title} {{An investigation of the
  quantum J1-J2-J3 model on the honeycomb lattice}},}\ }\href {\doibase
  10.1007/s100510170273} {\bibfield  {journal} {\bibinfo  {journal} {The
  European Physical Journal B}\ }\textbf {\bibinfo {volume} {20}},\ \bibinfo
  {pages} {241} (\bibinfo {year} {2001})}\BibitemShut {NoStop}%
\bibitem [{\citenamefont {Mulder}\ \emph {et~al.}(2010)\citenamefont {Mulder},
  \citenamefont {Ganesh}, \citenamefont {Capriotti},\ and\ \citenamefont
  {Paramekanti}}]{Mulder2010SpiralLattice}%
  \BibitemOpen
  \bibfield  {author} {\bibinfo {author} {\bibfnamefont {A.}~\bibnamefont
  {Mulder}}, \bibinfo {author} {\bibfnamefont {R.}~\bibnamefont {Ganesh}},
  \bibinfo {author} {\bibfnamefont {L.}~\bibnamefont {Capriotti}}, \ and\
  \bibinfo {author} {\bibfnamefont {A.}~\bibnamefont {Paramekanti}},\
  }\bibfield  {title} {\enquote {\bibinfo {title} {{Spiral order by disorder
  and lattice nematic order in a frustrated Heisenberg antiferromagnet on the
  honeycomb lattice}},}\ }\href {\doibase 10.1103/PhysRevB.81.214419}
  {\bibfield  {journal} {\bibinfo  {journal} {Physical Review B}\ }\textbf
  {\bibinfo {volume} {81}},\ \bibinfo {pages} {214419} (\bibinfo {year}
  {2010})}\BibitemShut {NoStop}%
\bibitem [{\citenamefont {Bishop}\ \emph {et~al.}(2012)\citenamefont {Bishop},
  \citenamefont {Li}, \citenamefont {Farnell},\ and\ \citenamefont
  {Campbell}}]{Bishop2012TheModel}%
  \BibitemOpen
  \bibfield  {author} {\bibinfo {author} {\bibfnamefont {R.~F.}\ \bibnamefont
  {Bishop}}, \bibinfo {author} {\bibfnamefont {P.~H.}\ \bibnamefont {Li}},
  \bibinfo {author} {\bibfnamefont {D.~J.}\ \bibnamefont {Farnell}}, \ and\
  \bibinfo {author} {\bibfnamefont {C.~E.}\ \bibnamefont {Campbell}},\
  }\bibfield  {title} {\enquote {\bibinfo {title} {{The frustrated Heisenberg
  antiferromagnet on the honeycomb lattice: J 1-J 2 model}},}\ }\href {\doibase
  10.1088/0953-8984/24/23/236002} {\bibfield  {journal} {\bibinfo  {journal}
  {Journal of Physics Condensed Matter}\ }\textbf {\bibinfo {volume} {24}}
  (\bibinfo {year} {2012}),\ 10.1088/0953-8984/24/23/236002}\BibitemShut
  {NoStop}%
\bibitem [{\citenamefont {Bishop}\ \emph {et~al.}(2015)\citenamefont {Bishop},
  \citenamefont {Li}, \citenamefont {G{\"{o}}tze}, \citenamefont {Richter},\
  and\ \citenamefont {Campbell}}]{Bishop2015FrustratedParameters}%
  \BibitemOpen
  \bibfield  {author} {\bibinfo {author} {\bibfnamefont {R.~F.}\ \bibnamefont
  {Bishop}}, \bibinfo {author} {\bibfnamefont {P.~H.~Y.}\ \bibnamefont {Li}},
  \bibinfo {author} {\bibfnamefont {O.}~\bibnamefont {G{\"{o}}tze}}, \bibinfo
  {author} {\bibfnamefont {J.}~\bibnamefont {Richter}}, \ and\ \bibinfo
  {author} {\bibfnamefont {C.~E.}\ \bibnamefont {Campbell}},\ }\bibfield
  {title} {\enquote {\bibinfo {title} {{Frustrated Heisenberg antiferromagnet
  on the honeycomb lattice: Spin gap and low-energy parameters}},}\ }\href
  {\doibase 10.1103/PhysRevB.92.224434} {\bibfield  {journal} {\bibinfo
  {journal} {Physical Review B}\ }\textbf {\bibinfo {volume} {92}},\ \bibinfo
  {pages} {224434} (\bibinfo {year} {2015})}\BibitemShut {NoStop}%
\bibitem [{\citenamefont {Doi}\ \emph {et~al.}(2004)\citenamefont {Doi},
  \citenamefont {Hinatsu},\ and\ \citenamefont
  {Ohoyama}}]{Doi2004StructuralNi}%
  \BibitemOpen
  \bibfield  {author} {\bibinfo {author} {\bibfnamefont {Y.}~\bibnamefont
  {Doi}}, \bibinfo {author} {\bibfnamefont {Y.}~\bibnamefont {Hinatsu}}, \ and\
  \bibinfo {author} {\bibfnamefont {K.}~\bibnamefont {Ohoyama}},\ }\bibfield
  {title} {\enquote {\bibinfo {title} {{Structural and magnetic properties of
  pseudo-two-dimensional triangular antiferromagnets Ba3MSb2O9 (M = Mn, Co,
  and Ni)}},}\ }\href {\doibase 10.1088/0953-8984/16/49/009} {\bibfield
  {journal} {\bibinfo  {journal} {Journal of Physics: Condensed Matter}\
  }\textbf {\bibinfo {volume} {16}},\ \bibinfo {pages} {8923} (\bibinfo {year}
  {2004})}\BibitemShut {NoStop}%
\bibitem [{\citenamefont {Shirata}\ \emph {et~al.}(2012)\citenamefont
  {Shirata}, \citenamefont {Tanaka}, \citenamefont {Matsuo},\ and\
  \citenamefont {Kindo}}]{Shirata2012ExperimentalAntiferromagnet}%
  \BibitemOpen
  \bibfield  {author} {\bibinfo {author} {\bibfnamefont {Y.}~\bibnamefont
  {Shirata}}, \bibinfo {author} {\bibfnamefont {H.}~\bibnamefont {Tanaka}},
  \bibinfo {author} {\bibfnamefont {A.}~\bibnamefont {Matsuo}}, \ and\ \bibinfo
  {author} {\bibfnamefont {K.}~\bibnamefont {Kindo}},\ }\bibfield  {title}
  {\enquote {\bibinfo {title} {{Experimental Realization of a Spin-1/2
  Triangular-Lattice Heisenberg Antiferromagnet}},}\ }\href {\doibase
  10.1103/PhysRevLett.108.057205} {\bibfield  {journal} {\bibinfo  {journal}
  {Physical Review Letters}\ }\textbf {\bibinfo {volume} {108}},\ \bibinfo
  {pages} {057205} (\bibinfo {year} {2012})}\BibitemShut {NoStop}%
\bibitem [{\citenamefont {Zhou}\ \emph {et~al.}(2012)\citenamefont {Zhou},
  \citenamefont {Xu}, \citenamefont {Hallas}, \citenamefont {Silverstein},
  \citenamefont {Wiebe}, \citenamefont {Umegaki}, \citenamefont {Yan},
  \citenamefont {Murphy}, \citenamefont {Park}, \citenamefont {Qiu},
  \citenamefont {Copley}, \citenamefont {Gardner},\ and\ \citenamefont
  {Takano}}]{Zhou2012SuccessiveBa3CoSb2O9}%
  \BibitemOpen
  \bibfield  {author} {\bibinfo {author} {\bibfnamefont {H.~D.}\ \bibnamefont
  {Zhou}}, \bibinfo {author} {\bibfnamefont {C.}~\bibnamefont {Xu}}, \bibinfo
  {author} {\bibfnamefont {A.~M.}\ \bibnamefont {Hallas}}, \bibinfo {author}
  {\bibfnamefont {H.~J.}\ \bibnamefont {Silverstein}}, \bibinfo {author}
  {\bibfnamefont {C.~R.}\ \bibnamefont {Wiebe}}, \bibinfo {author}
  {\bibfnamefont {I.}~\bibnamefont {Umegaki}}, \bibinfo {author} {\bibfnamefont
  {J.~Q.}\ \bibnamefont {Yan}}, \bibinfo {author} {\bibfnamefont {T.~P.}\
  \bibnamefont {Murphy}}, \bibinfo {author} {\bibfnamefont {J.-H.}\
  \bibnamefont {Park}}, \bibinfo {author} {\bibfnamefont {Y.}~\bibnamefont
  {Qiu}}, \bibinfo {author} {\bibfnamefont {J.~R.~D.}\ \bibnamefont {Copley}},
  \bibinfo {author} {\bibfnamefont {J.~S.}\ \bibnamefont {Gardner}}, \ and\
  \bibinfo {author} {\bibfnamefont {Y.}~\bibnamefont {Takano}},\ }\bibfield
  {title} {\enquote {\bibinfo {title} {{Successive Phase Transitions and
  Extended Spin-Excitation Continuum in the S=1/2 Triangular-Lattice
  Antiferromagnet Ba3CoSb2O9}},}\ }\href {\doibase
  10.1103/PhysRevLett.109.267206} {\bibfield  {journal} {\bibinfo  {journal}
  {Physical Review Letters}\ }\textbf {\bibinfo {volume} {109}},\ \bibinfo
  {pages} {267206} (\bibinfo {year} {2012})}\BibitemShut {NoStop}%
\bibitem [{\citenamefont {Susuki}\ \emph {et~al.}(2013)\citenamefont {Susuki},
  \citenamefont {Kurita}, \citenamefont {Tanaka}, \citenamefont {Nojiri},
  \citenamefont {Matsuo}, \citenamefont {Kindo},\ and\ \citenamefont
  {Tanaka}}]{Susuki2013MagnetizationBa3CoSb2O9}%
  \BibitemOpen
  \bibfield  {author} {\bibinfo {author} {\bibfnamefont {T.}~\bibnamefont
  {Susuki}}, \bibinfo {author} {\bibfnamefont {N.}~\bibnamefont {Kurita}},
  \bibinfo {author} {\bibfnamefont {T.}~\bibnamefont {Tanaka}}, \bibinfo
  {author} {\bibfnamefont {H.}~\bibnamefont {Nojiri}}, \bibinfo {author}
  {\bibfnamefont {A.}~\bibnamefont {Matsuo}}, \bibinfo {author} {\bibfnamefont
  {K.}~\bibnamefont {Kindo}}, \ and\ \bibinfo {author} {\bibfnamefont
  {H.}~\bibnamefont {Tanaka}},\ }\bibfield  {title} {\enquote {\bibinfo {title}
  {{Magnetization Process and Collective Excitations in the S=1/2
  Triangular-Lattice Heisenberg Antiferromagnet Ba3CoSb2O9}},}\ }\href
  {\doibase 10.1103/PhysRevLett.110.267201} {\bibfield  {journal} {\bibinfo
  {journal} {Physical Review Letters}\ }\textbf {\bibinfo {volume} {110}},\
  \bibinfo {pages} {267201} (\bibinfo {year} {2013})}\BibitemShut {NoStop}%
\bibitem [{\citenamefont {Quirion}\ \emph {et~al.}(2015)\citenamefont
  {Quirion}, \citenamefont {Lapointe-Major}, \citenamefont {Poirier},
  \citenamefont {Quilliam}, \citenamefont {Dun},\ and\ \citenamefont
  {Zhou}}]{Quirion2015MagneticMeasurements}%
  \BibitemOpen
  \bibfield  {author} {\bibinfo {author} {\bibfnamefont {G.}~\bibnamefont
  {Quirion}}, \bibinfo {author} {\bibfnamefont {M.}~\bibnamefont
  {Lapointe-Major}}, \bibinfo {author} {\bibfnamefont {M.}~\bibnamefont
  {Poirier}}, \bibinfo {author} {\bibfnamefont {J.~A.}\ \bibnamefont
  {Quilliam}}, \bibinfo {author} {\bibfnamefont {Z.~L.}\ \bibnamefont {Dun}}, \
  and\ \bibinfo {author} {\bibfnamefont {H.~D.}\ \bibnamefont {Zhou}},\
  }\bibfield  {title} {\enquote {\bibinfo {title} {{Magnetic phase diagram of
  Ba3CoSb2O9 as determined by ultrasound velocity measurements}},}\ }\href
  {\doibase 10.1103/PhysRevB.92.014414} {\bibfield  {journal} {\bibinfo
  {journal} {Physical Review B}\ }\textbf {\bibinfo {volume} {92}},\ \bibinfo
  {pages} {014414} (\bibinfo {year} {2015})}\BibitemShut {NoStop}%
\bibitem [{\citenamefont {Ma}\ \emph {et~al.}(2016)\citenamefont {Ma},
  \citenamefont {Kamiya}, \citenamefont {Hong}, \citenamefont {Cao},
  \citenamefont {Ehlers}, \citenamefont {Tian}, \citenamefont {Batista},
  \citenamefont {Dun}, \citenamefont {Zhou},\ and\ \citenamefont
  {Matsuda}}]{Ma2016StaticBa3CoSb2O9}%
  \BibitemOpen
  \bibfield  {author} {\bibinfo {author} {\bibfnamefont {J.}~\bibnamefont
  {Ma}}, \bibinfo {author} {\bibfnamefont {Y.}~\bibnamefont {Kamiya}}, \bibinfo
  {author} {\bibfnamefont {T.}~\bibnamefont {Hong}}, \bibinfo {author}
  {\bibfnamefont {H.}~\bibnamefont {Cao}}, \bibinfo {author} {\bibfnamefont
  {G.}~\bibnamefont {Ehlers}}, \bibinfo {author} {\bibfnamefont
  {W.}~\bibnamefont {Tian}}, \bibinfo {author} {\bibfnamefont {C.}~\bibnamefont
  {Batista}}, \bibinfo {author} {\bibfnamefont {Z.}~\bibnamefont {Dun}},
  \bibinfo {author} {\bibfnamefont {H.}~\bibnamefont {Zhou}}, \ and\ \bibinfo
  {author} {\bibfnamefont {M.}~\bibnamefont {Matsuda}},\ }\bibfield  {title}
  {\enquote {\bibinfo {title} {{Static and Dynamical Properties of the Spin-1/2
  Equilateral Triangular-Lattice Antiferromagnet Ba3CoSb2O9}},}\ }\href
  {\doibase 10.1103/PhysRevLett.116.087201} {\bibfield  {journal} {\bibinfo
  {journal} {Physical Review Letters}\ }\textbf {\bibinfo {volume} {116}},\
  \bibinfo {pages} {087201} (\bibinfo {year} {2016})}\BibitemShut {NoStop}%
\bibitem [{\citenamefont {Maksimov}\ \emph {et~al.}(2016)\citenamefont
  {Maksimov}, \citenamefont {Zhitomirsky},\ and\ \citenamefont
  {Chernyshev}}]{Maksimov2016Field-inducedAntiferromagnets}%
  \BibitemOpen
  \bibfield  {author} {\bibinfo {author} {\bibfnamefont {P.~A.}\ \bibnamefont
  {Maksimov}}, \bibinfo {author} {\bibfnamefont {M.~E.}\ \bibnamefont
  {Zhitomirsky}}, \ and\ \bibinfo {author} {\bibfnamefont {A.~L.}\ \bibnamefont
  {Chernyshev}},\ }\bibfield  {title} {\enquote {\bibinfo {title}
  {{Field-induced decays in XXZ triangular-lattice antiferromagnets}},}\ }\href
  {\doibase 10.1103/PhysRevB.94.140407} {\bibfield  {journal} {\bibinfo
  {journal} {Physical Review B}\ }\textbf {\bibinfo {volume} {94}},\ \bibinfo
  {pages} {140407} (\bibinfo {year} {2016})}\BibitemShut {NoStop}%
\bibitem [{\citenamefont {Li}\ \emph {et~al.}(2020{\natexlab{b}})\citenamefont
  {Li}, \citenamefont {Zhang},\ and\ \citenamefont
  {Zhang}}]{Li2020HighMonolayers}%
  \BibitemOpen
  \bibfield  {author} {\bibinfo {author} {\bibfnamefont {X.}~\bibnamefont
  {Li}}, \bibinfo {author} {\bibfnamefont {Z.}~\bibnamefont {Zhang}}, \ and\
  \bibinfo {author} {\bibfnamefont {H.}~\bibnamefont {Zhang}},\ }\bibfield
  {title} {\enquote {\bibinfo {title} {{High throughput study on magnetic
  ground states with Hubbard U corrections in transition metal dihalide
  monolayers}},}\ }\href {\doibase 10.1039/C9NA00588A} {\bibfield  {journal}
  {\bibinfo  {journal} {Nanoscale Advances}\ }\textbf {\bibinfo {volume} {2}},\
  \bibinfo {pages} {495} (\bibinfo {year} {2020}{\natexlab{b}})}\BibitemShut
  {NoStop}%
\bibitem [{\citenamefont {Moreo}\ \emph {et~al.}(1990)\citenamefont {Moreo},
  \citenamefont {Dagotto}, \citenamefont {Jolicoeur},\ and\ \citenamefont
  {Riera}}]{Moreo1990IncommensurateModels}%
  \BibitemOpen
  \bibfield  {author} {\bibinfo {author} {\bibfnamefont {A.}~\bibnamefont
  {Moreo}}, \bibinfo {author} {\bibfnamefont {E.}~\bibnamefont {Dagotto}},
  \bibinfo {author} {\bibfnamefont {T.}~\bibnamefont {Jolicoeur}}, \ and\
  \bibinfo {author} {\bibfnamefont {J.}~\bibnamefont {Riera}},\ }\bibfield
  {title} {\enquote {\bibinfo {title} {{Incommensurate correlations in the t-J
  and frustrated spin-1/2 Heisenberg models}},}\ }\href {\doibase
  10.1103/PhysRevB.42.6283} {\bibfield  {journal} {\bibinfo  {journal}
  {Physical Review B}\ }\textbf {\bibinfo {volume} {42}},\ \bibinfo {pages}
  {6283} (\bibinfo {year} {1990})}\BibitemShut {NoStop}%
\bibitem [{\citenamefont
  {Chubukov}(1991)}]{Chubukov1991First-orderAntiferromagnets}%
  \BibitemOpen
  \bibfield  {author} {\bibinfo {author} {\bibfnamefont {A.}~\bibnamefont
  {Chubukov}},\ }\bibfield  {title} {\enquote {\bibinfo {title} {{First-order
  transition in frustrated quantum antiferromagnets}},}\ }\href {\doibase
  10.1103/PhysRevB.44.392} {\bibfield  {journal} {\bibinfo  {journal} {Physical
  Review B}\ }\textbf {\bibinfo {volume} {44}},\ \bibinfo {pages} {392}
  (\bibinfo {year} {1991})}\BibitemShut {NoStop}%
\bibitem [{\citenamefont {Rastelli}\ and\ \citenamefont
  {Tassi}(1992)}]{Rastelli1992NonlinearPhase}%
  \BibitemOpen
  \bibfield  {author} {\bibinfo {author} {\bibfnamefont {E.}~\bibnamefont
  {Rastelli}}\ and\ \bibinfo {author} {\bibfnamefont {A.}~\bibnamefont
  {Tassi}},\ }\bibfield  {title} {\enquote {\bibinfo {title} {{Nonlinear
  effects in the spin-liquid phase}},}\ }\href {\doibase
  10.1103/PhysRevB.46.10793} {\bibfield  {journal} {\bibinfo  {journal}
  {Physical Review B}\ }\textbf {\bibinfo {volume} {46}},\ \bibinfo {pages}
  {10793} (\bibinfo {year} {1992})}\BibitemShut {NoStop}%
\bibitem [{\citenamefont {Ferrer}(1993)}]{Ferrer1993Spin-liquidLattice}%
  \BibitemOpen
  \bibfield  {author} {\bibinfo {author} {\bibfnamefont {J.}~\bibnamefont
  {Ferrer}},\ }\bibfield  {title} {\enquote {\bibinfo {title} {{Spin-liquid
  phase for the frustrated quantum Heisenberg antiferromagnet on a square
  lattice}},}\ }\href {\doibase 10.1103/PhysRevB.47.8769} {\bibfield  {journal}
  {\bibinfo  {journal} {Physical Review B}\ }\textbf {\bibinfo {volume} {47}},\
  \bibinfo {pages} {8769} (\bibinfo {year} {1993})}\BibitemShut {NoStop}%
\bibitem [{\citenamefont {Ceccatto}\ \emph {et~al.}(1993)\citenamefont
  {Ceccatto}, \citenamefont {Gazza},\ and\ \citenamefont
  {Trumper}}]{Ceccatto1993NonclassicalAntiferromagnets}%
  \BibitemOpen
  \bibfield  {author} {\bibinfo {author} {\bibfnamefont {H.~A.}\ \bibnamefont
  {Ceccatto}}, \bibinfo {author} {\bibfnamefont {C.~J.}\ \bibnamefont {Gazza}},
  \ and\ \bibinfo {author} {\bibfnamefont {A.~E.}\ \bibnamefont {Trumper}},\
  }\bibfield  {title} {\enquote {\bibinfo {title} {{Nonclassical disordered
  phase in the strong quantum limit of frustrated antiferromagnets}},}\ }\href
  {\doibase 10.1103/PhysRevB.47.12329} {\bibfield  {journal} {\bibinfo
  {journal} {Physical Review B}\ }\textbf {\bibinfo {volume} {47}},\ \bibinfo
  {pages} {12329} (\bibinfo {year} {1993})}\BibitemShut {NoStop}%
\bibitem [{\citenamefont {Reuther}\ \emph {et~al.}(2011)\citenamefont
  {Reuther}, \citenamefont {W{\"{o}}lfle}, \citenamefont {Darradi},
  \citenamefont {Brenig}, \citenamefont {Arlego},\ and\ \citenamefont
  {Richter}}]{Reuther2011QuantumModel}%
  \BibitemOpen
  \bibfield  {author} {\bibinfo {author} {\bibfnamefont {J.}~\bibnamefont
  {Reuther}}, \bibinfo {author} {\bibfnamefont {P.}~\bibnamefont
  {W{\"{o}}lfle}}, \bibinfo {author} {\bibfnamefont {R.}~\bibnamefont
  {Darradi}}, \bibinfo {author} {\bibfnamefont {W.}~\bibnamefont {Brenig}},
  \bibinfo {author} {\bibfnamefont {M.}~\bibnamefont {Arlego}}, \ and\ \bibinfo
  {author} {\bibfnamefont {J.}~\bibnamefont {Richter}},\ }\bibfield  {title}
  {\enquote {\bibinfo {title} {{Quantum phases of the planar antiferromagnetic
  J1−J2−J3 Heisenberg model}},}\ }\href {\doibase
  10.1103/PhysRevB.83.064416} {\bibfield  {journal} {\bibinfo  {journal}
  {Physical Review B}\ }\textbf {\bibinfo {volume} {83}},\ \bibinfo {pages}
  {064416} (\bibinfo {year} {2011})}\BibitemShut {NoStop}%
\end{thebibliography}%

\end{document}